\documentclass[aps,pra,preprint, superscriptaddress,showpacs, amsfonts]{revtex4-1}

\usepackage{graphicx}
\usepackage{color}
\usepackage{hyperref}
\usepackage{bm}

\hypersetup{colorlinks=false}

\begin{document}

\title{Hidden  Modes in Open  Disordered Media: Analytical, Numerical, and Experimental Results. }

\author{Yury P. Bliokh}
\affiliation{Physics Department, Technion-Israel Institute of Technology, Haifa 32000,Israel}
\affiliation{Center for Emergent Matter Science (CEMS), RIKEN, Wako-shi, Saitama, 351-0198, Japan}

\author{Valentin Freilikher}
\affiliation{Jack and Pearl Resnick Institute of Advanced Technology, Department of Physics, Bar-Ilan University, Ramat-Gan 52900, Israel}
\affiliation{Center for Emergent Matter Science (CEMS), RIKEN, Wako-shi, Saitama, 351-0198, Japan}

\author{Z. Shi}
\affiliation{Department of Physics, Queens College of the City University of New York, Flushing, NY 11367l, USA}
\affiliation{The Graduate Center, CUNY, 365 Fifth Avenue, New York, NY 10016, USA}

\author{A. Z. Genack}
\affiliation{Department of Physics, Queens College of the City University of New York, Flushing, NY 11367l, USA}
\affiliation{The Graduate Center, CUNY, 365 Fifth Avenue, New York, NY 10016, USA}

\author{Franco Nori}
\affiliation{Center for Emergent Matter Science (CEMS), RIKEN, Wako-shi, Saitama, 351-0198, Japan}
\affiliation{Center for Theoretical Physics, Department of Physics, CSCS, University of Michigan, Ann Arbor, Michigan 48109-1040, USA}

\begin{abstract}

We explore numerically, analytically, and experimentally the relationship between quasi-normal modes (QNMs) and transmission resonance (TR) peaks  in the transmission spectrum of one-dimensional (1D) and quasi-1D open disordered systems. It is shown that for weak disorder there exist two types of the eigenstates: ordinary QNMs which are associated with a TR, and hidden QNMs which do not exhibit peaks in transmission or within the sample. The distinctive feature of the hidden modes  is that unlike ordinary ones, their lifetimes remain constant in a wide range of the strength of disorder. In this range,  the averaged ratio of the number   of transmission peaks $N_{\rm res}$ to the number   of QNMs $N_{\rm mod}$, $N_{\rm res}/N_{\rm mod}$, is insensitive to the type and degree of disorder and  is close to the value $\sqrt{2/5}$, which we derive analytically in the weak-scattering approximation. The physical nature of the hidden modes is illustrated in simple examples with a few scatterers. The analogy between ordinary and hidden QNMs and the segregation of superradiant states and trapped modes is discussed. When the coupling to the environment is tuned by an external edge reflectors, the superradiace transition is reproduced. 
Hidden modes have been also found in microwave measurements in quasi-1D open disordered samples. The microwave measurements and modal analysis of transmission in the crossover to localization in quasi-1D systems give a ratio of $N_{\rm res}/N_{\rm mod}$ close to $\sqrt{2/5}$. In diffusive quasi-1D samples, however, $N_{\rm res}/N_{\rm mod}$ falls as the effective number of transmission eigenchannels $M$ increases. Once $N_{\rm mod}$ is divided by $M$, however, the ratio $N_{\rm res}/N_{\rm mod}$ is close to the ratio found in 1D. 

\end{abstract}

\maketitle

\section{Introduction}

Two powerful perspectives have helped clarify the nature of wave propagation in open random systems. One of them, relates to the leakage of waves through the boundaries of the system and can be described in terms of quasi-normal modes (QNMs), which are the extension to open structures of the notion of normal modes in closed systems \cite{1,2,5,newJ,newM,newR,newS,newT,1974c}. 
The eigenfrequencies of the QNMs are complex, with imaginary parts that are the inverses of the lifetimes of the QNMs. The second perspective is that of transmission through random systems \cite{1977a,1979a,10a}.
For multichannel samples, transmission is  most conveniently described in terms of the transmission matrix, $t$, whose elements are field transmission coefficients \cite{1981c,1984a,1988c}. The transmittance is the sum of eigenvalues of the Hermitian matrix $tt^{\dagger }$. Some of these eigenvalues are close to unity even in weakly-transmitting samples \cite{1984a,1988c,Imry,1990g}.
Knowledge of the transmission matrix makes it possible to  manipulate the incident wavefront to enhance or suppress  total transmission through random media \cite{newA1,8a,2012g,12,22a} and to focus transmitted radiation at selected points \cite{newF}. The control over  transmitted radiation can be exploited to improve images washed out by random scattering and to facilitate the detection and location of objects \cite{newF}. 
The great potential of such algorithms for a host of practical applications has recently attracted attention in both the physics
\cite{newF}
and mathematics communities \cite{1} and references therein. 

In open regular homogeneous systems (e.g. single quantum potential wells, optical
cavities, or microwave resonators) each peak in transmission, or transmission resonance (TR),
is associated with a QNM (\cite{Moiseev} and references therein), so that the resonant frequency is close to the real part of the corresponding eigenvalue. 
However, despite extensive research and much recent progress the
connection between QNMs and TRs in disordered open systems still requires a better physical understanding and mathematical justification,

To this end, it is instructive to look for insights in 1D systems. It is
well-known \cite{1977a,Lifshitz} that the transmission of a long enough 1D
disordered system is typically exponentially small. At the same time, there
exists a set of frequencies at which the transmission coefficient has a
local maximum (peak in transmission), and some of these are close to unity \cite{Lifshitz,we2,we1}. In 1D, each peak is
associated with an eigenstate which is a solution of the wave equation
with outgoing boundary conditions (a pole of the S-matrix).

Quite surprisingly, much less  are studied the properties of QNMs in 1D systems with weak disorder where the localization length is smaller than the size of the sample.  In this paper we show 
for the first time that in completely open one-dimensional disordered systems, two different types of QNMs can exist: 
ordinary QNMs, associated 
with resonant
transmission peaks and  hidden QNMs unrelated to any maxima in the
transmission spectrum. The hidden modes exist due to random scattering and arise as soon as an arbitrarily small disorder is introduced.
The imaginary parts of
the eigenfrequencies of hidden QNMs vary with increasing disorder in an
unusual manner. Typically, stronger disorder leads to stronger
localization of modes with eigenfrequencies that approach the real axis.
However, the imaginary part of a hidden mode${}^\prime$s eigenfrequency, depending on the boundary conditions, either  is independent of strength of random scattering or even increases from the onset of disorder.
Surprisingly, the average ratio of the number of ordinary modes to the total
number of QNMs in a given frequency interval is independent of the type of
disorder and remains close to the constant $\sqrt{2/5}$ over wide ranges
of the strength of disorder and of the total length of the system. The value 
$\sqrt{2/5}$ follows from the general statistical properties of random
trigonometric polynomials \cite{Edelman}. As the scattering strength and/or the
length of the system increase, hidden QNMs eventually become ordinary.

The situation is different in multi-channel random systems in which a genuine diffusive regime exists. The degree of spectral overlap is expressed in the Thouless number, $\delta$, which gives the ratio of the typical width $\delta \nu$ and spacing $\Delta\nu$ of QNMs, $\delta=\delta\nu/\Delta\nu$ \cite{5,1974c,1977a}. The typical linewidth $\delta \nu$ is essentially equal to the field correlation frequency over which there is typically a single peak in the transmission spectrum. The density of peaks is therefore $1/\delta \nu$. On the other hand, the inverse level spacing $1/\Delta \nu$ is equal to the density of states (DOS) of the medium. Thus the ratio $N_{\rm res}/N_{\rm mod}$ can be expected to be close to $\Delta\nu/\delta\nu=1/\delta$ for diffusive waves. The localization threshold lies at $\delta=1$ \cite{1974c,1977a,1979a}; $\delta$ may be much larger than 1 for diffusive waves so that $N_{\rm res}/N_{\rm mod}\sim 1/\delta$  and may be small. 
For localized waves, the number of channels that contribute effectively to transmission, $M$, approached unity and transport becomes effectively one-dimensional \cite{M}. For example, the statistics of transmittance are then in accord with the single parameter scaling hypothesis \cite{Genack PNAS}. 
It is worth noticing that although the statistics of the eigenstates of disordered systems is a subject of intensive investigations for already more that two decades (see, for example \cite{RefA,RefB,RefC,RefD,RefE,RefF,RefG}), the statistics of the transmission resonances (peaks in transmission spectra) is much less studied. The comparison of these two is a challenging problem for future investigations. Here we find that 
a connection can be made between the present 1D calculations of $N_{\rm res}/N_{\rm mod}$ and measurements in multichannel diffusive systems. This is done by comparing ratio of $N_{\rm res}$ to the number of QNMs divided by $M$, $MN_{\rm res}/N_{\rm mod}$ in multichannel systems to the ratio $N_{\rm res}/N_{\rm mod}$ in 1D, where $M=1$.

\section{Quasi-normal modes of  open systems}
We first consider a generic 1D system composed of $N+1$ scatterers separated by $N$
intervals and attached to two semi-infinite leads.
The eigenfunctions $\psi_m (x,t)$ are solutions of the wave equation satisfying the outgoing boundary conditions, which
means that there are no right/left-propagating waves in the left/right lead. Each eigenfunction is a
superposition of two counter-propagating monochromatic waves $\psi
_{m}(x)^{(\pm )}e^{-i\omega _{m}t}$.  The eigenfunction in the $j$th
 layer, $\psi_{m,j}^{(\pm )}(x)$, is equal to $a_{m,j}^{(\pm )}e^{\pm ik_{m}x}$, and the amplitudes $a_{m,j}^{(\pm
)}$ in adjacent layers are connected by a transfer matrix. The wave numbers $k_{m}$ are complex-valued and form the discrete set  $k_{m}^{(\mathrm{mod)}}=k_{m}^{\prime }-ik_{m}^{\prime \prime }$, $%
k^{\prime \prime }_m>0$, so that the frequencies $\omega _{m}^{(\mathrm{mod)}}=ck_{m}^{(\mathrm{mod)}}.$
These eigenfunctions are QNMs. Note that all
distances hereafter are measured in optical lengths.

In what follows, the scatterers and the distances between them are
characterized by the reflection coefficients $r_{j}\equiv r_{0}+\delta r_{j}$
and  thicknesses $d_{j}\equiv d_{0}+\delta d_{j}$, respectively. The
random values $\delta r_{j}$ and $\delta d_{j}$ are distributed in certain
intervals, and $\sum\delta d_{j} =0$.  The last condition means that
the  length $L$ of the sample is equal to $Nd_{0}$. 

To explicitly introduce a variable strength $s$ of disorder, we replace all
reflection coefficients
by $sr_{j} $, and assume (unless otherwise
specified) that the coefficients $r_{j}$ are homogeneously distributed in
the interval $(-1,1)$.
This enables to keep track of the evolution of the QNM eigenvalues $k_{m}^{(\mathrm{mod)}}$
as the disorder increases.
The condition $\sum\delta d_{j}=0$ ensures that
any random realization with the same $N$ contains the same number of QNMs $N_{\rm mod}=\Delta kL/\pi$ in a given interval $\Delta k$ of the wavenumbers. 

At the beginning, let us consider the QNMs of a regular resonator of the length $L=Nd_0$ assuming that all reflection coefficients except $r_1=r_L$ and $r_{N+1}=r_R$ are equal to zero. 
In this case the real and imaginary parts of the QNM
eigenvalues $k_{m}^{(\mathrm{mod)}}$ are 
\begin{eqnarray}
k_{m}^{\prime } &=&\frac{1}{2L}\cdot \left\{ 
\begin{array}{r}
\pi +2\pi m,\hspace{5mm}\mathrm{when}\hspace{2mm}r_{L}r_{R}>0, \\ 
2\pi m,\hspace{5mm}\mathrm{when}\hspace{2mm}r_{L}r_{R}<0,%
\end{array}%
\right.  \label{eq1} \\
k_{m}^{\prime \prime } &=&-\frac{1}{2L}\ln |r_{L}r_{R}|,  \label{eq2}
\end{eqnarray}
where $m=0,1,2,\ldots$

\begin{figure}[t]
\centering \scalebox{0.4}{\includegraphics{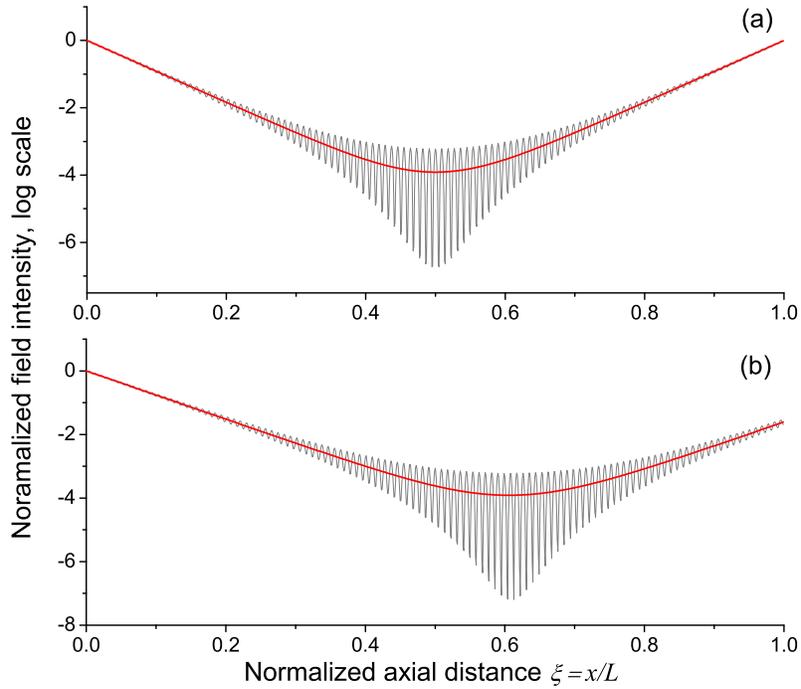}}
\caption{Spatial distribution of the intensity $I_m(x)$ of  quasi-normal mode (thin light gray curve) and $\bar{I}_m(x)$ 
(thick red curve) in a regular resonator with (a) symmetric ($|r_L|=|r_R|$) and (b) asymmetric ($|r_L|<|r_R|$) walls. \label{Fig0}}
\end{figure}

In what follows,   instead of
the intensity of the $m$th  mode,  $I_m(x)=\left|\psi_m^{(+)}(x)+\psi_m^{(-)}(x)\right|^2$, 
we consider the quantity  $\bar{I}_{m}(x)=|\psi _{m}^{(+)}(x)|^{2}+|\psi_{m}^{(-)}(x)|
^{2}$, which is $I_{m}(x)$  averaged over fast
oscillations caused by the interference of
the left- and right-propagating waves. Examples
of  these functions for 
resonators
are shown in Fig.~\ref{Fig0}a,b. 
There $\bar{I}_n(x)$ is distributed along the system as $\bar{I}_m(x)\propto
\cosh[2k^{\prime \prime}(x-x^\ast)]$,  where $x^{\ast }=L[1-\ln (|r_{R}/r_{L}|)/\ln(|r_{R}r_{L}|)]/2$.  When $|r_L|=|r_R|$, 
the minimum of the intensity is located at the center of the system, and in
an asymmetric case shifts to the boundary with  higher  reflection coefficient. This property will be used when analyzing the properties of the QNMs  of the disordered system.

\begin{figure}[tbh]
\centering \scalebox{0.3}{\includegraphics{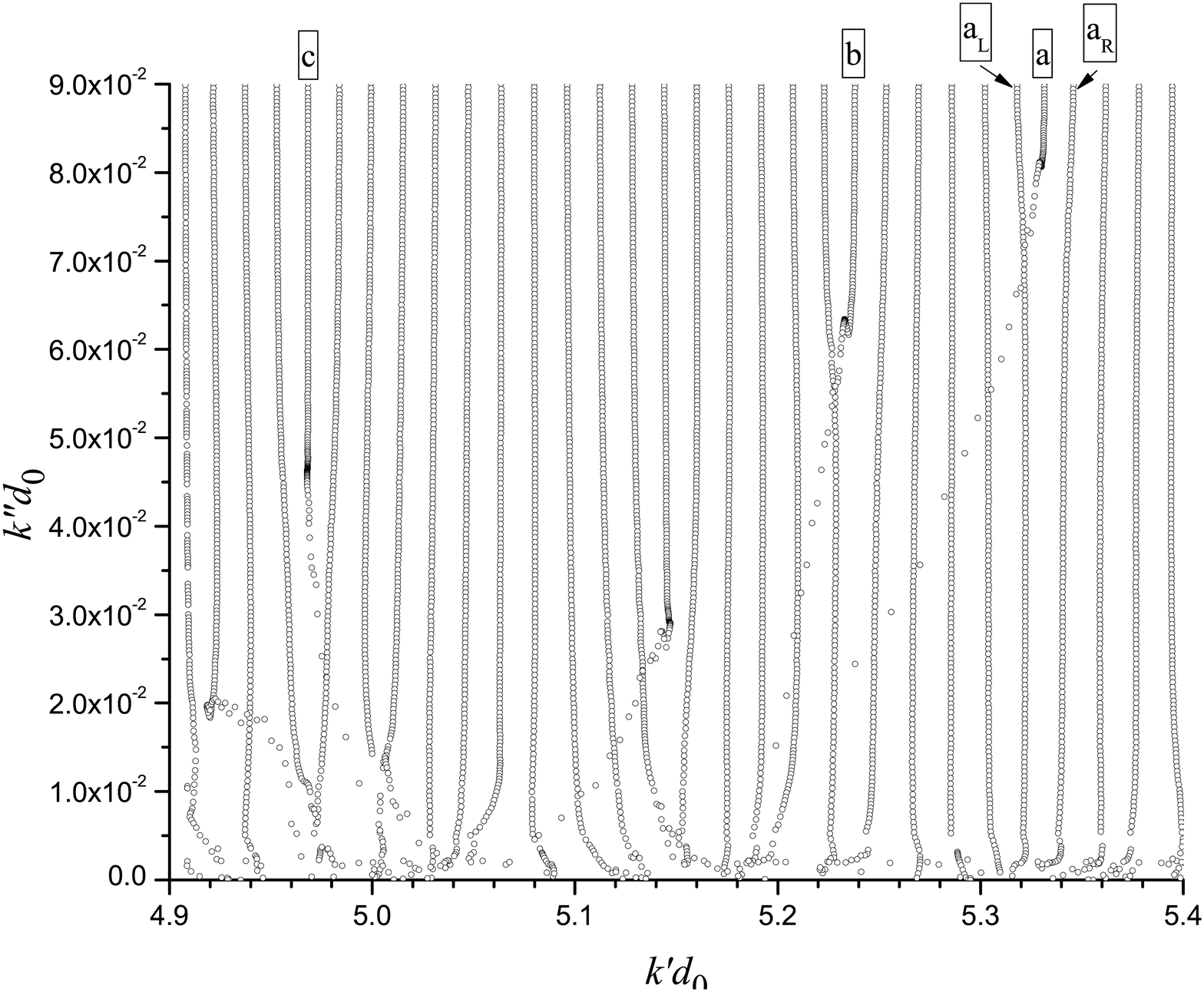}}
\caption{Motion of QNMs' eigenvalues under disorder strength growth. Eigenvalues move from above. Some eigenvalues (examples are marked by ``a'', ``b'', and``c'') stop their motion and ``stand still''.  }
\label{FigAA}
\end{figure}

\begin{figure}[tbh]
\centering \scalebox{0.3}{\includegraphics{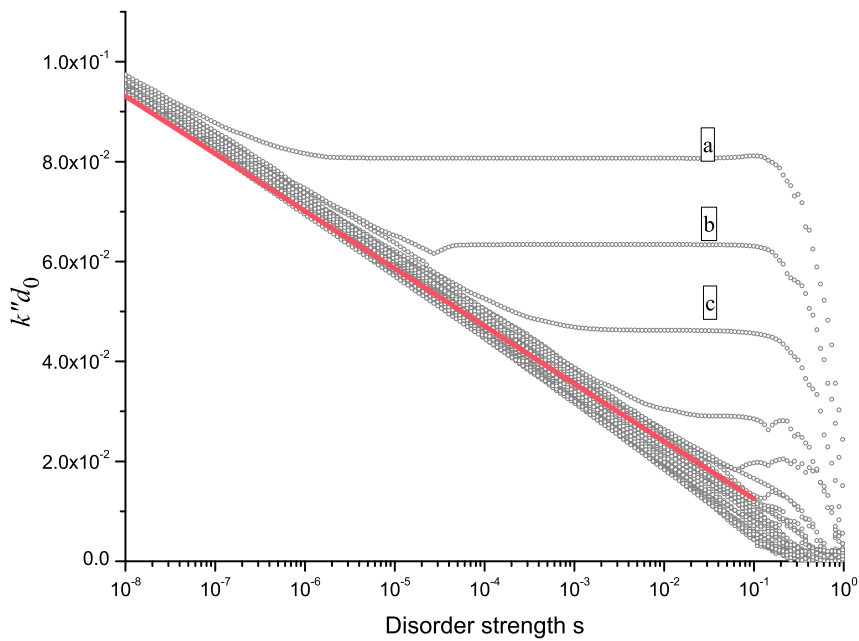}}
\caption{Variation of $k^{\prime\prime}(s)$. Some $k^{\prime\prime}$ (marked by the same letters as in Fig~\ref{FigAA}) are independent of $s$ in a broad range (several order of magnitude!) of $s$ variation. Note that length of the system is equal or larger than localization length when $s\geq 0.1$. Red line shows the dependence $k^{\prime\prime}(s)$, described by Eq.~(\ref{eqQ}) }
\label{FigB}
\end{figure}

In a disordered sample, the reflection coefficients $r_i$ are random and scaled by the parameter $s$. The evolution of the eigenvalues $k^{(\rm mod)}(s)$ as the parameter $s$ grows shows that QNMs separate into two essentially different types. 
There are ordinary QNMs whose lifetimes, defined by the value of  $1/k^{\prime\prime}$,
increase monotonically with $s$. Simultaneously, there are ``hidden'' QNMs (the origin of this term will be explained in the next section), whose lifetimes are substantially smaller than the lifetimes of ordinary QNMs and remain constant when $s$ varies over many orders of magnitude. 
Figures \ref{FigAA} and \ref{FigB} show trajectories of the QNMs' eigenvalues $k^{(\rm mod)}(s)=k^\prime(s)-ik^{\prime\prime}(s)$ as the parameter $s$ grows, and dependencies $k^{\prime\prime}(s)$. 

Our numerical calculations show that when external reflectors are added at the edges of the sample,  the imaginary parts of some of the hidden modes increase with the strength of disorder.

\begin{figure}[tbh]
\centering \scalebox{0.4}{\includegraphics{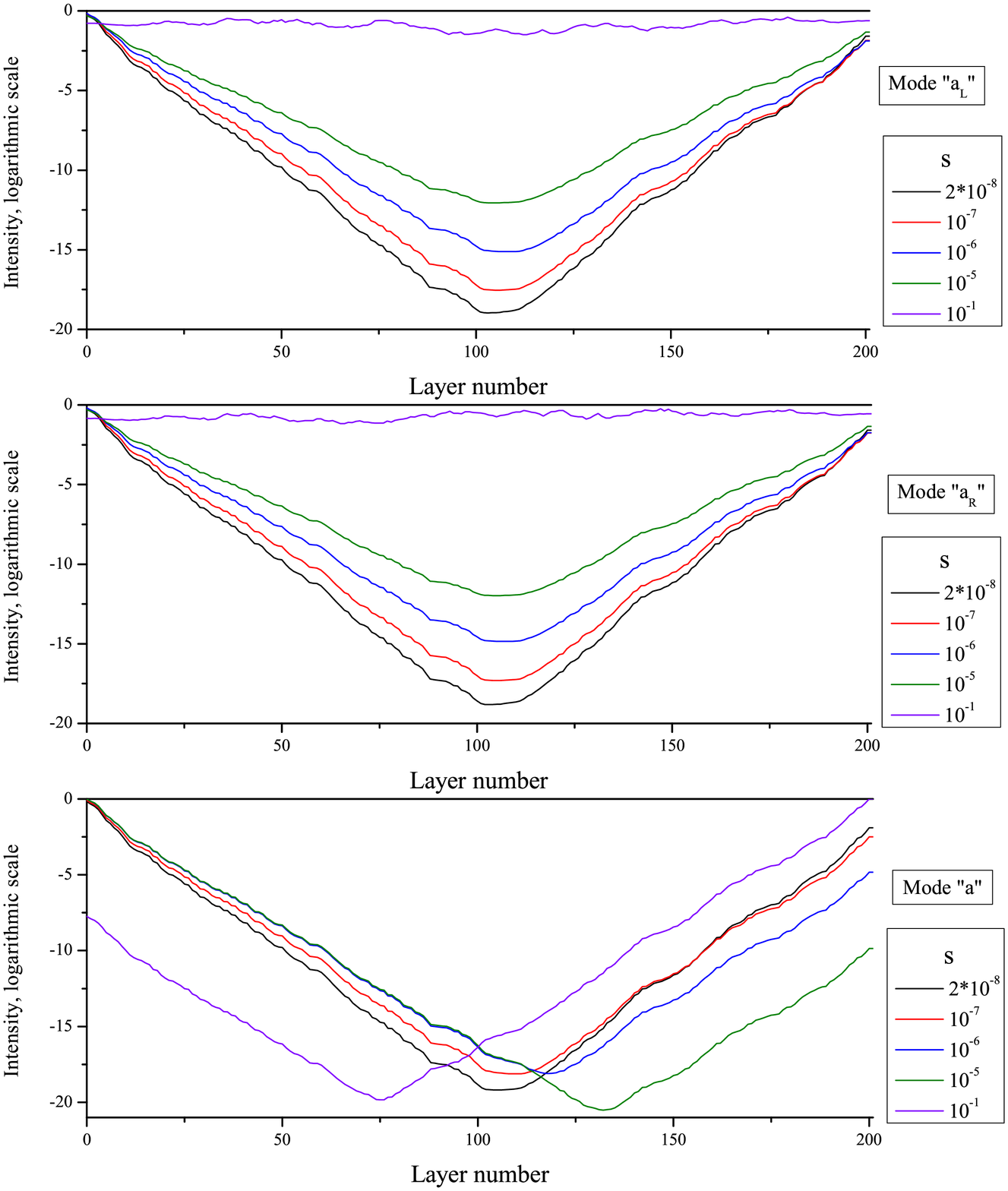}}
\caption{Intensity  distribution $\bar{I}(x)$ along the system for ordinary QNMs ``$a_L$'' (a)  and ``$a_R$'' (b), and hidden QNM ``$a$'' (c) for different values of the disorder strength $s$. Notations for the modes are the same as in Figs.~\ref{FigAA} and \ref{FigB}.}
\label{FNew1}
\end{figure}

The spatial distributions of the intensity $\bar{I}(x)$ along the system are also different for ordinary and hidden QNMs. The evolution of $\bar{I}(x)$ as the strength of the disorder $s$ grows is shown in Fig.~\ref{FNew1}. 

Initially, when $s$ is so small that values of $k^{\prime\prime}$ are almost equal for both types of modes, the distributions $\bar{I}(x)$ are practically identical and evolve in the same manner: the minimum is placed near the center of the sample, and slopes (which are $k^{\prime\prime}$) decrease as the disorder strength $s$ grows (see Fig.~\ref{FNew1}). These distributions are
similar to the distribution of the intensity in the regular resonator with a small imbalance between the reflection coefficients $r_L$ and $r_R$ of the resonator walls.

When $k^{\prime\prime}(s)$ of the hidden mode ``$a$'' reaches its plateau (see Fig.~ \ref{FigB}), the minimum of its distribution shifts from the center, as in the resonator with strong imbalance between the reflection coefficients $r_L$ and $r_R$. The slope of the distribution of hidden modes remains constant ($k^{\prime\prime}$ is independent of $s$ on the plateau), whereas the slopes of all ordinary modes are equal and continue to decrease as the parameter $s$ grows. 
The difference between the distributions of ordinary and hidden QNMs is that the ordinary modes are concentrated  near both edges of the system, while the hidden mode is nestled at one edge. 

It is important to stress that this separation of the QNMs into two types occurs when the disorder strength $s$ is small so that the localization length $\ell_{\rm loc}$ is large relative to the system length $L$, $\ell_{\rm loc}\gg L$. Thus, this phenomenon is not related to Anderson localization, but, as it will be shown below, manifests itself also when $\ell_{\rm loc}\ll L$.

Notwithstanding that at $s\to 1$, the lifetimes of all hidden modes increase, these modes are much more resistant to disorder: they become localized at far stronger disorder than ordinary states. As can be seen, for example, in Fig.~\ref{FigB}, at $s\sim 0.4-0.5$, when $L\sim 15-20\ell_{\rm loc}$, the difference between the imaginary parts of ordinary and some of hidden QNMs is about  of one order of magnitude. 

\section{Transmission resonances in 1D systems}

We now consider the transmission of an incident wave through the system. The  wavenumbers at which the transmission coefficient reaches its local maximum and the corresponding fields inside the system are  transmission resonances (TR). The QNMs and TRs are interrelated. In what follows, we explore  the relation between QNMs and TRs, in particular,  study the differences between the spectra of TRs and QNMs.

It is easy to show that in a resonator, the wave numbers $k_{m}^{(\mathrm{res})}$ of the transmission resonances coincide with the real parts $k_{m}^{\prime }$ given by Eq.~(\ref{eq1}), and  there is a one-to-one correspondence between QNMs and TRs so that the number of resonances $N_{\mathrm{res}}$ is equal to
the number of QNMs, $N_{\mathrm{mod}}$, in a given frequency interval. The same relation also exists in periodic systems (periodic sets $r_{j}$ and $d_{j}$) \cite{Settimi}.

\begin{figure}[tbh]
\centering \scalebox{0.8}{\includegraphics{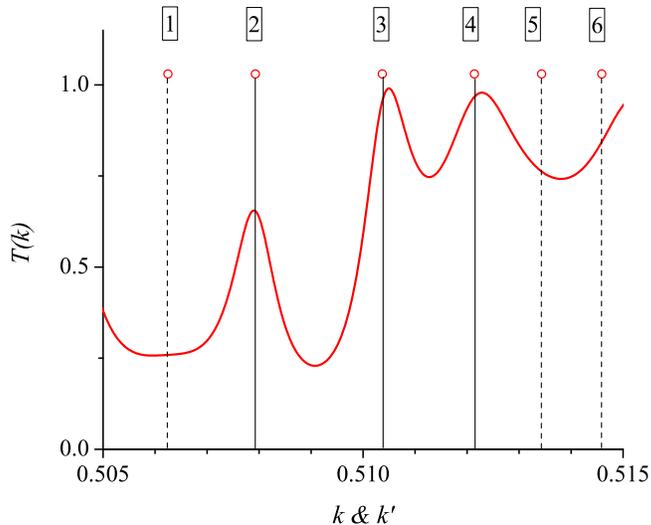}}
\caption{Transmission spectrum $T(k)$ at $s=0.1$ ($L\simeq \ell_{\rm loc}$). The black
solid (dashed) vertical lines indicate the $k_{n}^{\prime }$ values of the hidden (ordinary) QNMs. Every maximum in the transmission spectrum can be associated with ordinary QNM (\# 2,3,4), whereas  hidden QNMs (\#1,5,6)  are not associated with the transmission resonances. }
\label{QM_TR}
\end{figure}

\begin{figure}[tbh]
\centering \scalebox{0.6}{\includegraphics{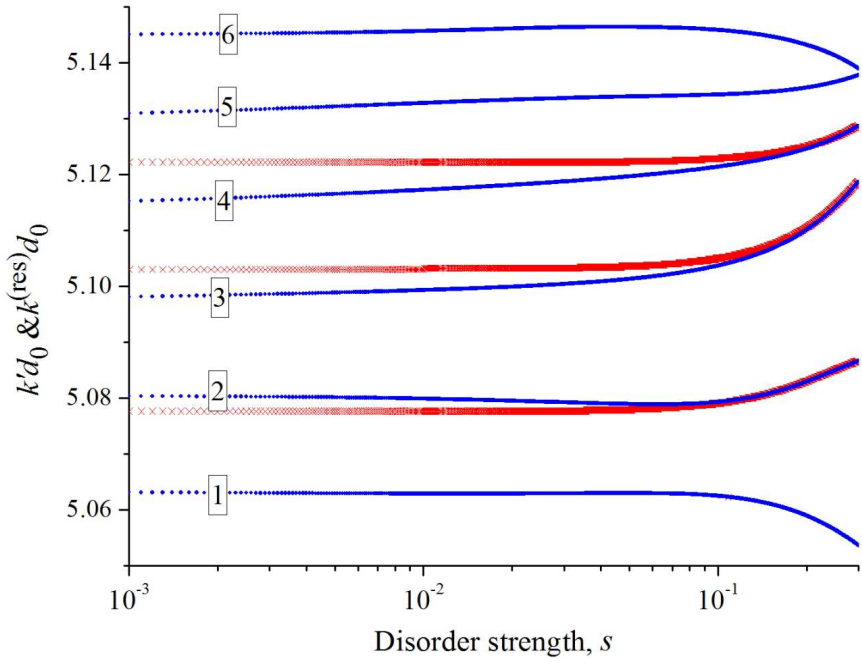}}
\caption{Variation of the wave numbers $k^{(\mathrm{res})}(s)$ (red crosses) and $k^{\prime }(s)$ (closed blue circles) with the strength $s$ of the disorder. QNMs are numbered as
in Fig.~\ref{QM_TR}. It is seen that for ordinary QNMs, $k^{(\mathrm{%
res})}(s)$ and $k^{\prime }(s)$ practically coincide, whereas there are no resonances associated with hidden QNMs (\#1,5,6). }
\label{QM_TR_s}
\end{figure}

In disordered systems, the relation between QNMs and TRs is quite different.  While each TR has its partner among the QNMs, 
the reverse is not true: there are hidden QNMs that are not associated with any maximum in transmission as shown in Figs.~\ref{QM_TR} and \ref{QM_TR_s} 

\begin{figure}[h]
\centering \scalebox{0.7}{\includegraphics{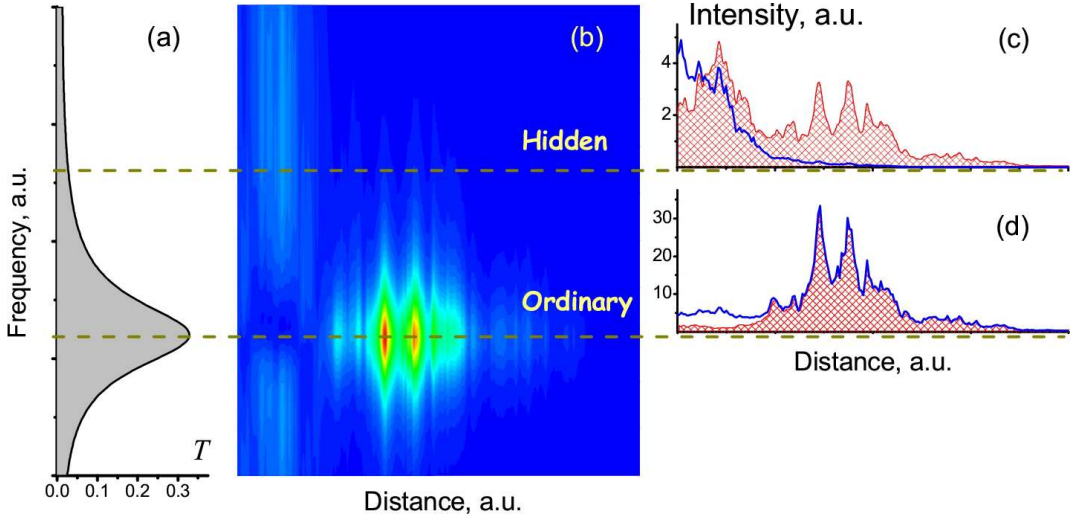}}
\caption{ Difference between ordinary and hidden QNMs. (a)  transmittance spectrum. (b)  distribution of the incident wave intensity into the sample as a function of frequency and distance.  There are two QNMs with nearby real parts of eigenfrequencies ${\rm Re}\,\omega^{\rm(mod)}$, marked by dashed  lines in the panels (a) and (b). Distributions of the intensities of eigenfunctions of hidden and ordinary QNMs along the sample are shown by thick blue lines in panels (c) and (d), correspondingly. Hatched red areas in panels (c) and (d) show intensity distributions of the incident waves whose frequencies coincide with ${\rm Re}\,\omega^{\rm(mod)}$ of hidden  and ordinary QNMs, corresondingly.         }
\label{FigIntro}
\end{figure}

Figure~\ref{FigIntro} illustrates another fundamental difference between the ordinary and hidden QNMs. The  ordinary QNMs whose real parts of the complex-valued eigenfrequency, ${\rm Re}\,\omega^{\rm(mod)}$ lie in a given frequency interval, can be determined from the transmittance spectrum $T(\omega)$ of 1D disordered samples, because each peak in the spectrum  corresponds to a frequency whose value $\omega^{(\rm res)}$ practically coincides with ${\rm Re}\,\omega^{\rm(mod)}$. Moreover, when disorder is strong enough, so that $L>\ell_{\rm loc}$, the distribution of the transmitted wave intensity along the sample reconstructs very closely the shape of the intensity of ordinary QNM eigenfunctions. In contrast, a hidden QNM is invisible (this explains the origin of the term "hidden") in the transmittance spectrum and its intensity distribution is indistinguishable from that at a non-resonant frequency.

Note that although the hidden modes are not displayed in the amplitude of the transmission coefficient, they are manifested in the phase of the transmission coefficient. The density of states at a frequency $\omega $ is proportional to the derivative with respect to frequency of the phase of the complex transmission coefficient \cite{avish}. Our numerical calculations show that each hidden mode adds  $\pi$  to the total phase shift of the transmission coefficient  exactly in the same way as  ordinary QNMs.

\begin{figure}[tbh]
\centering \scalebox{1.0}{\includegraphics{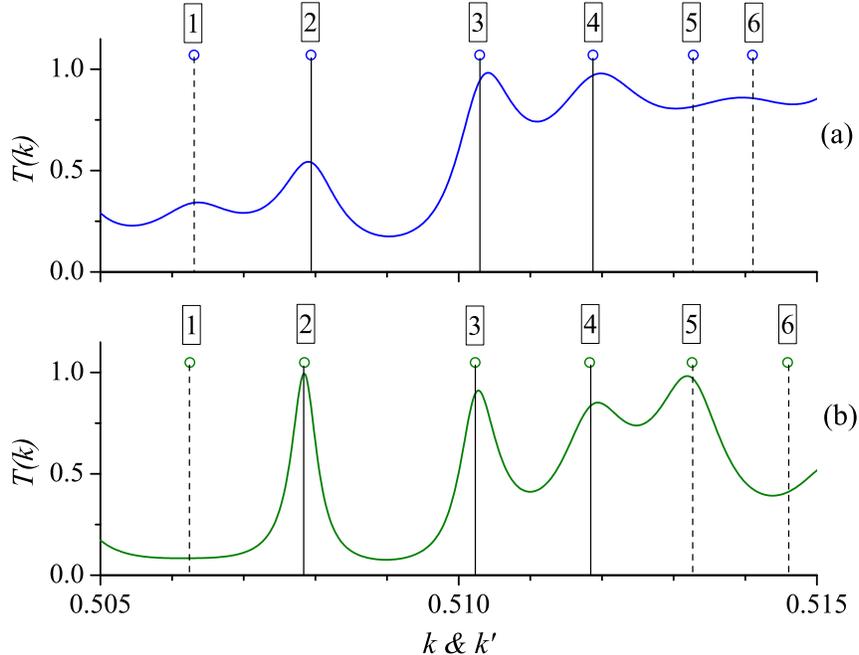}}
\caption{Hidden modes \#1,5, and 6, which are shown in Fig.~\ref{QM_TR}, can be transformed in the ordinary ones by additional end reflections. (a) -- additional right-end reflection   transforms the hidden modes \#1 and 6 into the ordinary modes; mode \#5 remains invisible in the transmission spectrum. (b) -- in contrast, the left-end reflection transforms mode \#5 into a ordinary mode, while modes \#1 and 6 remain invisible. }
\label{Fig4a}
\end{figure}

The evolution of a hidden QNM as the degree of disorder grows is analogous
to the evolution of a mode in a regular resonator when one of its edges
becomes less transparent. This means that a hidden mode may be transformed
into an ordinary (i.e., made visible in the transmission) by increasing the
reflectivity of the corresponding edge of the sample, as illustrated in Fig.~\ref{Fig4a}.

The sample, whose transmission spectrum is shown in Fig.~\ref{QM_TR}, contains three hidden QNMs (\# 1,5, and 6) in the given spectral range. Distributions of the intensity $\bar{I}(n)$ ($n$ is the layer number) for QNMs \#1 and 6 are similar to the distributions in resonators with right reflection coefficient $r_R$ smaller than the left reflection coefficient $r_L$, $r_R/r_L<1$. The intensity distribution of QNM \#5 is characterized by the opposite inequality $r_R/r_L>1$. 
When the value of the right-end reflection coefficient is increased,  new resonances appear in transmission for the initially hidden modes \#1 and 6, while mode \#5 remains hidden (Fig.~\ref{Fig4a}a). 
In contrast, increasing the left-end
reflection coefficient transforms QNM \#5 into a ordinary mode, whereas
QNMs \#1 and 6 remain hidden in the transmission spectrum (Fig.~\ref{Fig4a}b).

Important to stress that the separation of QNMs into two types, ordinary and hidden, occurs already at a very small disorder strength, $s\to 0$, when the localization length is larger than the sample length, $\ell_{\rm loc}\gg L$.
  
The ensemble-averaged of the ratio of the number of transmission resonances, $N_{\mathrm{res}}$, which is the number of ordinary modes, to
the total number of QNMs, $N_{\mathrm{mod}}$, has been calculated
numerically for a variety of randomly layered samples with different types of
disorder (random reflection coefficients of the layers, $r_{j}$,
and/or random thicknesses $d_{j}$, with rectangular and Gaussian
distribution functions) in broad ranges of the disorder strength $s$ and of
the length of the realizations $N$.

\begin{figure}[tbh]
\centering \scalebox{0.5}{\includegraphics{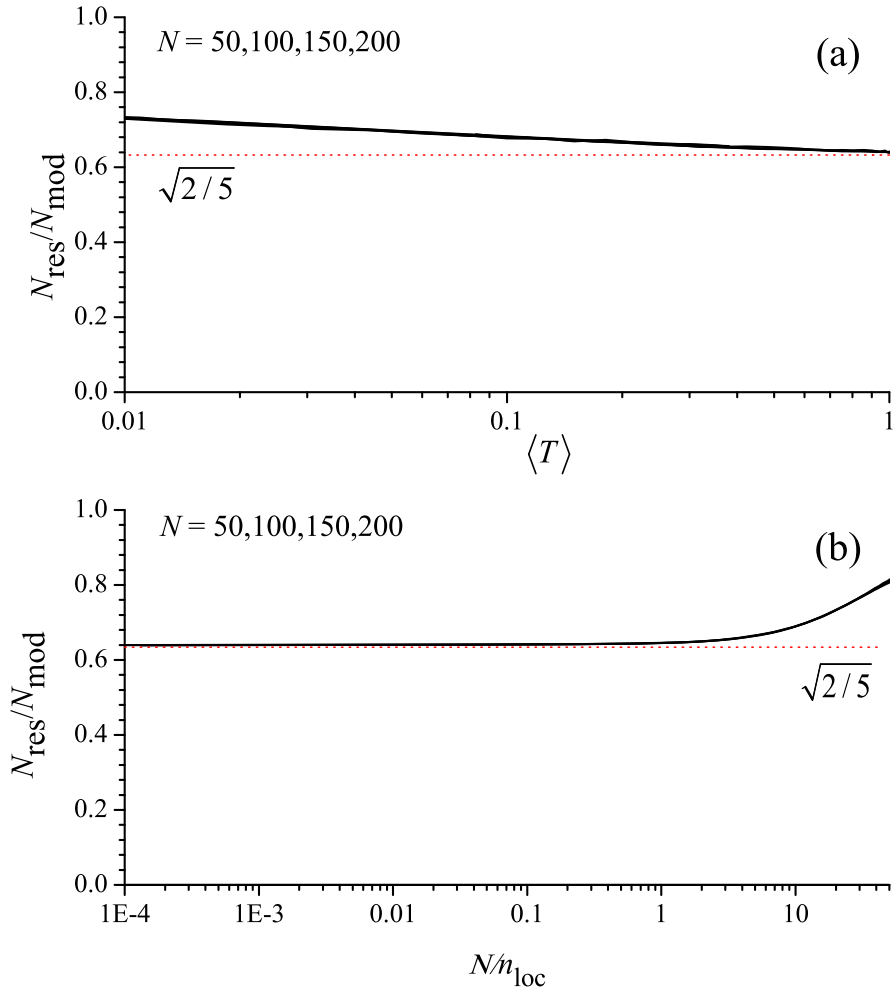}}
\caption{Ratio $N_{\rm res}/N_{\rm mod}$ as a function of the ensemble-averaged transmission coefficient $\langle T\rangle$ [panel(a)], and as a function of ratio of $N$ to the localization length $n_{\mathrm{loc}}$
[panel (b)]  for systems of various lengths (number of layers $N$). The horizontal dashed red line marks  $\sqrt{2/5}$. } 
\label{Fig7a}
\end{figure}

Figure~\ref{Fig7a} shows the average of $N_{\mathrm{res}}/N_{\mathrm{mod}}$ over $10^{4}$
random realizations  as a function of the ensemble-averaged transmission coefficient $\langle T\rangle$ [panel (a)], and as a function of ratio of $N$ to the localization length $n_{\mathrm{loc}}$
(measured in numbers of layers), $N/n_{\rm loc}$ [panel (b)] for samples with  $N=50,100,150$, and  $200$ layers. At this scaling, all functions $N_{\mathrm{res}}(\langle T\rangle)/N_{\mathrm{mod}}$ and $N_{\mathrm{res}}(N/n_{\rm loc})/N_{\mathrm{mod}}$  for
samples of different lengths merge in a single  curve.

It is seen in Fig.~\ref{Fig7a}  that the difference between $N_{\mathrm{res}}$ and $N_{\mathrm{mod}}$ appears  when $n_{\mathrm{loc}}\gg N$, and the  ratio $N_{\mathrm{res}}/N_{\mathrm{mod}}$  varies weakly even when $n_{\mathrm{loc}}\ll N$. Moreover, independently of the samples parameters, the average ratio $N_{\mathrm{res}}/N_{\mathrm{mod}}$ tends to the constant $\sqrt{2/5}$ when $n_{\mathrm{loc}}\to\infty$. Thus, the existence of hidden modes and the universality of their relative number is a general feature of 1D disordered systems not specifically related to localization. 

\section{Measurements of transmission eigenchannels and transmission resonances in multichannel systems}

It is of interest to explore the ratio of the numbers of local maxima in transmission and QNMs in random \textit{multichannel} systems and to compare to results for 1D systems. We consider quasi-one dimensional (quasi-1D) samples with reflecting sides and transverse dimensions $W$ much smaller than the sample length $L$ and  localization length $\ell_{\rm loc}=N_{\rm chan}\ell$, $W<\ell_{\rm loc},L$. Here, $N_{\rm chan}$ is the number of channels or freely-propagating transverse modes in the perfectly conducting leads or empty waveguide leading to the sample and $\ell$ is the transport mean free path. The incident channels are thoroughly mixed by scattering within the sample. In contrast to transmission in 1D samples with a single transmission channel, transmission through quasi-1D samples is described by the  field transmission matrix $t$ with elements $t_{ba}$ between all $N_{\rm chan}$ incident and outgoing channels, $a$ and $b$, respectively. From the transmission matrix, we may distinguish three types of transmission variables in quasi-1D samples: the intensity $T_{ba}=|t_{ba}|^2$, the total transmission, ${T}_{a}=\sum_{b=1}^{N_{\rm chan}} { T}_{ba}$, and transmittance, 
$T=\sum_{a,b=1}^{N_{\rm chan}}\left|t_{ba}\right|^2$.
The transmittance is analogous to the electronic conductance in units of the quantum of conductance $e^2/h$ \cite{Landauer,10a,1981c}. The ensemble average value of the transmittance $T$ is equal to the dimensionless conductance, $g=\langle T\rangle$, which characterizes the crossover from diffusive to localized waves. In diffusive samples, the dimensionless conductance is equal to the Thouless number, $g=\delta$ and the localization threshold is reached when $g=\delta=1$ \cite{1977a, 1979a}.

Significant differences between results in 1D and quasi-1D geometries can be expected since propagation can be diffusive in quasi-1D samples with length greater than the mean free path but smaller than the localization length, $\ell<L<\ell_{\rm loc}=N_{\rm chan}\ell$, whereas a diffusive regime does not exist in 1D since $\ell_{\rm loc}=\ell$ \cite{Abr}. For diffusive waves, QNMs overlap spectrally and may coalesce into a single peak in the transmittance spectrum. Thus we might expect that the QNMs within a typical linewidth form a single peak in transmission so that the ratio $N_{\rm res}/N_{\rm mod}$ is the ratio  of the mode spacing to the mode linewidth. The mode linewidth is related to the correlation frequency in the transmission spectra,
but the mode spacing cannot be readily ascertained once modes overlap. 

The transmittance can also be expressed as $T=\sum_{n=1}^{N_{\rm chan}} \tau_n$, where the $\tau_n$ are the eigenvalues of the matrix product $tt^\dagger$ \cite{1981c}. The transmission matrix provides a basis for comparison between results for 1D and quasi-1D, which is often more direct than a comparison based on QNMs, since the statistics of the contribution of different modes to transmission is not well-established, whereas the contribution of different channels is simply the sum of the transmission eigenvalues. In addition, transmission eigenchannels are orthogonal, whereas the waveform in transmission  for spectrally-adjacent modes are strongly correlated \cite{5}  so that the transmission involves interference between modes. 

The transmission eigenvalue may be obtained from the singular value decomposition of the transmission matrix, $t=U\Lambda V^\dagger$ \cite{Been}. Here, $U$ and $V$ are unitary matrices and $\Lambda$ is a diagonal matrix with elements $\sqrt{\tau_n}$. The incident fields of the eigenchannels on the incident surface, $v_n$, which are the columns of $V$, in the singular-value decomposition are orthogonal, as are the corresponding outgoing eigenchannels, $u_n$. Only a fraction of the $N_{\rm chan}$ eigenchannels contribute appreciably to the transmission \cite{1984a}. In diffusive samples, the transmission is dominated by $g$ channels with $\tau_{n}>1/e$ \cite{3, Imry}, while a single eigenchannel dominates transmission for localized samples. The statistics of transmission depend directly on the participation number of transmission eigenhannels, $M\equiv (\sum_{n=1}^{N_{\rm chan}} \tau_n)^2/\sum_{n=1}^{N_{\rm chan}} \tau_n^2$ \cite{M}. $M$ is equal to $3g/2$ \cite{M} for diffusive waves and approaches unity in the localized limit \cite{M,Genack PNAS}.

\subsection{Numerical simulations}

To explore the ratio $N_{\rm res}/N_{\rm mod}$ over a broad range of $g=\langle T \rangle$ for multichannel disordered waveguides in the crossover from diffusive to localized waves, we carry out numerical simulations for a scalar wave propagating through a two dimensional disordered waveguide with reflecting sides and semi-infinite leads. 
For diffusive samples in which there is considerable mode overlap since $\delta=\delta \nu/\Delta \nu >1$ \cite{1977a}, ($\delta\nu$ and $\Delta\nu$ are the linewidth and the distance between spectral lines) the density of states (DOS), and from this the number of QNMs within the spectrum, can be obtained from the sum of the derivatives of the composite phase  of the transmission eigenchannel \cite{2015a}. The derivative of the composite phase of the $n${th} eigenchannel  is equal to the dwell time of the photon within the sample in the eigenchannel. The total number of modes $N_{\rm mod}$ in a given frequency interval is then the integral over this interval of the DOS. This has allowed us to determine the ratio  $N_{\rm res}/N_{\rm mod}$  in the crossover to localization.

\begin{figure}[tbh]
\centering \scalebox{0.9}{\includegraphics{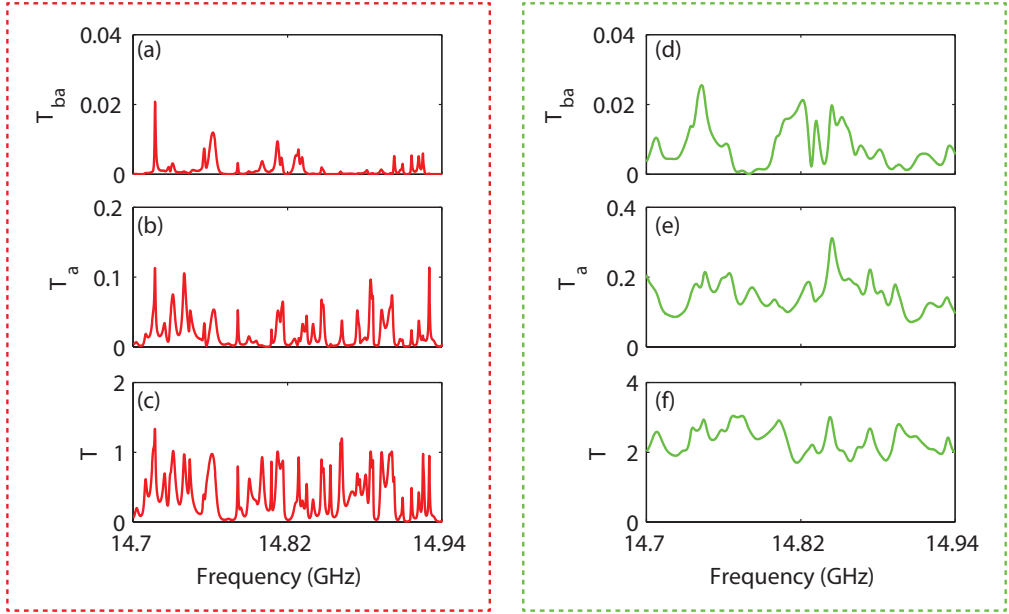}}
\caption{ Spectra of intensity, total transmission and transmittance for a localized sample drawn from a random ensemble with {\it g} = 0.3 (a)-(c) and a diffusive sample taken from an ensemble with {\it g} = 2.1 (d)-(f). Sharper spectral features are observed and spatial averaging is seen to be less effective in smoothing the spectra for localized waves than for diffusive waves.}\label{AAA}
\end{figure}

Simulations are carried out by discretizing  the wave equation 
\begin{equation}\label{Azi2}
\nabla^2 E(x,y) + k_0^2\epsilon(x,y)E(x,y) = 0
\end{equation}
on a square grid and solved via the recursive Green function method \cite{stone91}. Here, $k_0$ is the wave vector in the leads. Also, $\epsilon(x,y)= 1\pm \delta\epsilon(x,y)$ is the spatially-varying dielectric function in the disordered region with $\delta\epsilon(x,y)$ chosen from a rectangular distribution and $\epsilon = 1$ in the empty leads. Reflections at the sample boundaries are minimal because the sample is index matched to its surroundings.
The product of $k_0$ at 14.7 GHz and the grid spacing is set to unity. In the frequency range studied, the leads attached to the random waveguide support $N_{\rm chan}=16$ channels which are the propagating waveguide modes. In our scalar quasi-1D simulations for a sample with a width $W$, the number of channels at frequencies above the cutoff frequency is the integer part of $2W/\lambda$. These channels should not be confused with the QNMs of the random medium which correspond to resonances of the medium with Lorentzian lines centered at distinct frequencies. In the simulations, the length of the sample $L$ is equal to 500 in units of the grid spacing except for one deeply localized sample with $g=0.12$, for which $L$ = 800 and the width of the sample $W$ is $16\pi$. Typical spectra of intensity, total transmission and transmittance are shown in Fig.~\ref{AAA} for a diffusive sample with $g=2.1$ and for a localized sample with $g=0.3$. 

\begin{figure}[tbh]
\centering \scalebox{0.45}{\includegraphics{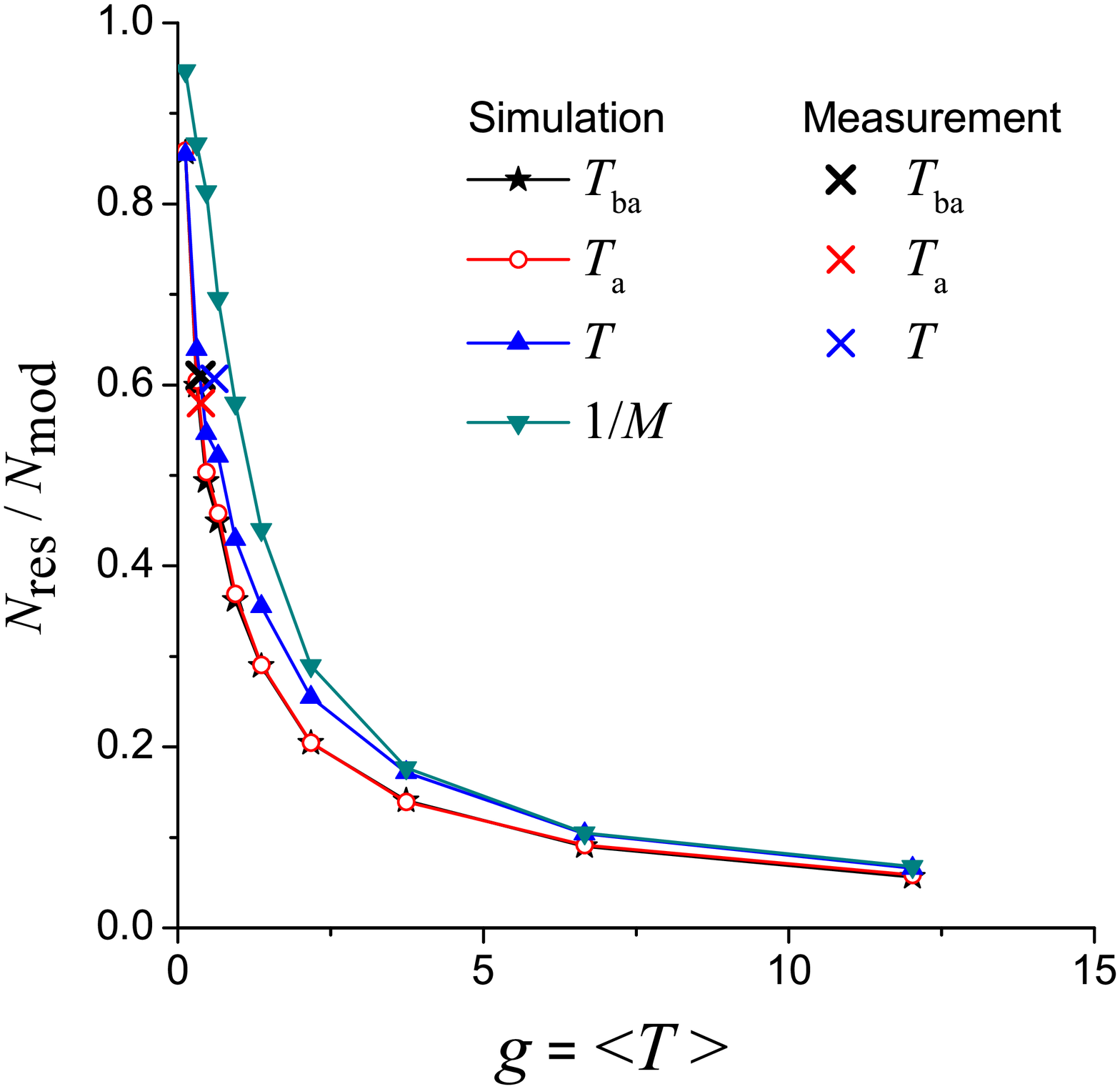}}
\caption{ Variation of the $N_{\rm res}/N_{\rm mod}$ for transmission, total transmission and transmittance vs. $g = \langle T\rangle$ for multichannel random samples in simulations. The ratios obtained from microwave measurements of spectra of the three transmission variables in a multichannel localized sample with  $g=0.37$ are shown as the cross symbols and are in good agreement with the simulations with a similar value of $g$. The value of $1/M$ found in the simulations is shown as overturned triangles.}\label{BBB}
\end{figure}

We find that the numbers of peaks in the spectra of intensity, total transmission and transmittance in a single sample are nearly the same for each of the samples shown in Fig.~\ref{AAA}. This is seen to be the case over a wide range of $\langle T \rangle$ in Fig.~\ref{BBB}. 

The DOS and so the number of QNMs within the spectrum in the samples of the same size are not affected by the strength of disorder so that the decreasing ratio $N_{\rm res}/N_{\rm mod}$ with increasing $\langle T \rangle$ reflects only the decreasing number of peaks in the transmission spectra due to the broadening of the modes and the consequent increase in their spectral overlap. Since there are typically $\delta$ QNMs within the mode linewidth for diffusive waves, we might expect the ratio $ N_{\rm res}/N_{\rm mod}$ to fall inversely with $M$, $N_{\rm res}/N_{\rm mod} \sim 1/\delta \sim 1/g \sim 3/2M$. For deeply  localized waves, however, this ratio is expected to approach unity as $M$ approaches unity. This suggests that $ N_{\rm res}/N_{\rm mod} \sim 1/M$. in this limit. A plot of $1/M$ in Fig.~\ref{BBB} shows that towards the diffusive and localized limits $1/M$ is close to the ratio $ N_{\rm res}/N_{\rm mod} $. 
For diffusive waves, the intensity correlation frequency does not change as the width of the sample changes for fixed length and scattering strength since it is tied to the time of the flight distribution, which is independent of $W$ \cite{Drake}. Since $N_{\rm res}$ is essentially the  width of the spectrum divided by the correlation frequency of the intensity, the number of peaks within the intensity spectrum does not change. However, $g$ and the DOS are proportional to $N_{\rm chan}$, so that $M$ increases with sample width and $ N_{\rm res}/N_{\rm mod} $ is inversely proportional to $M$. In addition, the propagation in a multichannel disordered sample is essentially 1D, when $M$ is approaching unity \cite{Genack PNAS}. 

\begin{figure}[tbh]
\centering \scalebox{0.53}{\includegraphics{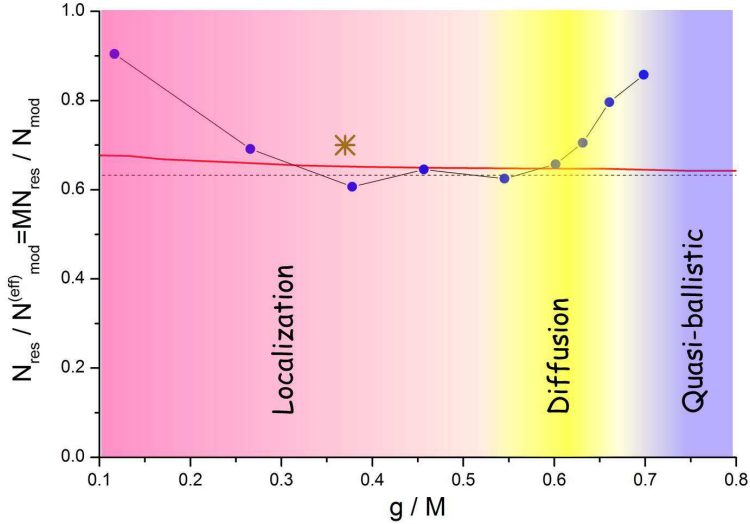}}
\caption{ Number of peaks in the transmission spectra per effective transmission eigenchannel,  $ N_{\rm res}/(N_{\rm mod}/M)$, is plotted as a function of $T=g/M$. The quantity $g/M$ is the effective transmission coefficient per effective transmission eigenvalue of the quasi-1D system. Such normalization of the conductance $g$ in quasi-1D samples makes possible a comparison with 1D systems. The red line corresponds to a 1D system; the experimental data is shown by the asterisk; the blue dots show the results of numerical simulations; and dashed line is drawn at the level $\sqrt{2/5}$. Beyond the diffusive regime the ratio plotted rises towards unity for ballistic propagation. For ballistic waves, each of the $N_{\rm channel}$ transmission eigenvalue is unity so that the transmittance is $N_{\rm channel}$ and all eigenchannels contribute equally to the transmittance so that $M=N_{\rm channel}$, yielding $g/M=1$.}\label{CCC}
\end{figure}

These results suggest that a comparison can be made between propagation in both 1D and multichannel systems via the ratio of the number of peaks in the transmission spectra to the number of modes normalized by $M$, $ N_{\rm res}/(N_{\rm mod}/M)$. This ratio may be expected to be close to unity for $L\gg\ell_{\rm loc}$. We consider the variation with $g=\langle T \rangle$ of the ratio $MN_{\rm res}/N_{\rm mod}$ in quasi-1D and compare this with the corresponding ratio in 1D in which $M=1$. The values of this ratio  in quasi-1D and 1D are  close, as seen in Fig.~\ref{CCC}.

\subsection{Microwave experiment}

For quasi-1D samples in the crossover to localization in which spectral overlap is moderate, it is possible to analyze the  measured field spectra to obtain the central frequencies of the QNMs and to compare these to peaks in transmission.  Spectral measurements of the transmittance $T$ were made in a copper waveguide of diameter 7.3 cm and of length 40 cm containing randomly positioned alumina spheres with index 3.14, over a random ensemble for which $g = 0.37$. The empty waveguide supports $N_{\rm chan}\sim 30$ propagation channels in the frequency range of the experiment: 10.0-10.24 GHz. The transmission matrix is determined from measurements of the field transmission coefficient between points on grids of 49 locations for the source antenna and detection antennas on the input and output surfaces of the waveguide for a single polarization with a grid spacing of 9 mm.
Such measurements of the transmission matrix in real space for a single polarization are incomplete. The distribution of transmission eigenvalues determined from these measurements may differ from theoretical calculations \cite{Been,2013j}. We find, however, that  the impact of incompleteness upon the statistics of transmittance and transmission eigenvalues is small as long as the number of measured channels is much greater than $M$, as is the case in these measurements of transmission in localized samples \cite{Genack PNAS}. Here $M=1.23$ and therefore the statistics of transmission are not affected by the incompleteness of the measurement \cite{Genack PNAS}. The influence of absorption in these samples is statistically removed by compensating for the enhanced decay of the field due to absorption \cite{1993a}. Different random sample configurations are obtained by briefly rotating and vibrating the sample tube. The probability distribution of the transmittance is in good agreement with the distribution calculated for this value of $g$ \cite{Mutt, Gopar, Froufe, Genack PNAS}.

We find the central frequencies and linewidths  of the QNMs within the frequency range of the measurements by carrying out a modal decomposition of the transmitted field. A given polarization component of the field  can be expressed as a sum of the contributions from each of the QNMs: 
\begin{equation}
E({\bf r},\omega)=\Sigma_m a_m({\bf r})\frac{\Gamma_m/2}{\Gamma_m/2+i\left(\omega-\omega_m\right)}.\label{Azi1}
\end{equation}
Here  $ a_m({\bf r})$ are complex-valued amplitudes of QNMs.  

\begin{figure}[tbh]
\centering \scalebox{0.5}{\includegraphics{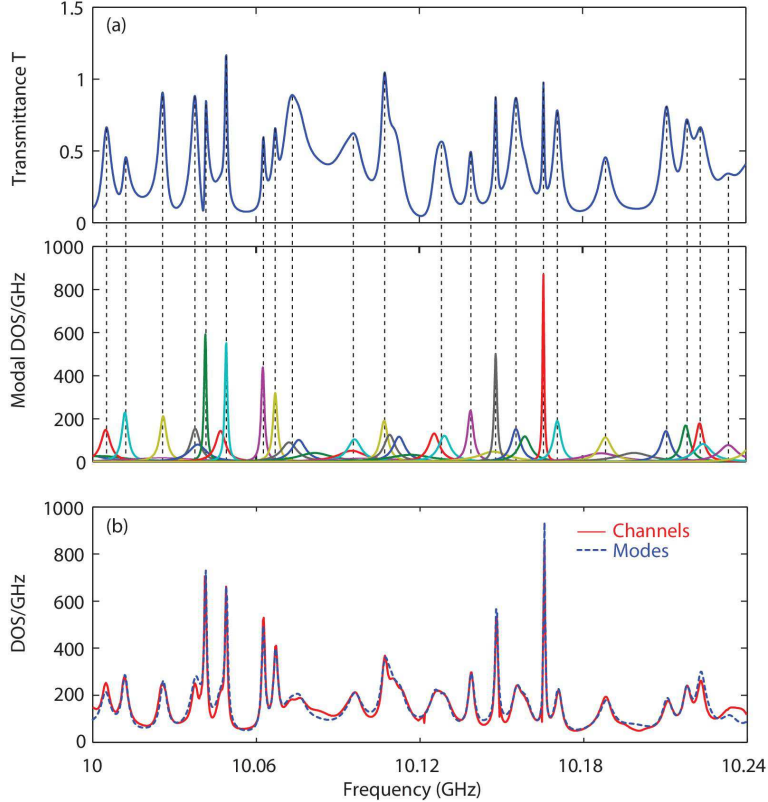}}
\caption{(a) Spectrum of transmittance $T$ and the individual modes. The integration of  each Lorentzian curve in the lower panel over the frequency yields the density of state of unity. There are 22 local maxima in the spectrum of $T$ 
and the number of modes are 39. (b) Spectrum of the density of states. The sum of all the Lorentzian curves above gives the DOS of the sample, 
which is seen to be in good agreement with the DOS [panel (b)] obtained via the summation of the composite phase derivatives of each transmission eigenchannel.
\label{DDD}}
\end{figure}

The central frequencies $\omega_m$ and linewidths $\Gamma_m$ of the modes are found by simultaneously fitting 45 field spectra. The transmittance as well as the Lorentzian lines for each QNM normalized to unity and the DOS, which is the sum of such Lorentzian lines over all QNMs 
are shown in Fig.~\ref{DDD} for a single random configuration. The DOS curves for different modes are plotted in different colors so that they can be distinguished more clearly. The DOS is also determined from the sum of the spectral derivatives of the composite phase of each transmission eigenchannel and plotted in Fig.~\ref{DDD}. The DOS determined from analyses of the QNMs and of the transmission eigenchannels are seen to be in agreement. 
The dashed vertical lines in Fig.~\ref{DDD} are drawn from the peaks in the transmittance spectra in (a) to the frequency axis in (b). As found in 1D simulations, each peak in $T$ is close to the frequency of a QNM, but many QNMs do not correspond to a distinct peak in the transmittance. Frequently, more than one QNM falls within a single peak in $T$. 

The ratio of the number of peaks in spectra of transmittance to the number of QNMs  averaged over a random ensemble of 40 configurations is 0.61, with a standard deviation of 0.057. This is indicated by the cross in Fig.~\ref{BBB} and is consistent with values of the ratio found in computer simulations. This value of this ratio is slightly smaller than the value 0.65 found in simulations for 1D sample with $\langle T \rangle=0.37$ found in 1D simulations, as seen in Fig.~\ref{Fig7a}. This may be attributed to the value  $M=1.23$  being larger than the value of unity in  1D. This reflects the tendency of the ratio to decrease with increasing $M$ as found for diffusive waves.

Equation (\ref{Azi1}) offers an explanation for the fact that the number of transmission resonances can be smaller than that of all QNMs. If, for example, the transmitted field is a sum of two modes, from Eq.~(\ref{Azi1}) it follows that the number of peaks in the transmission spectrum is either one or two, depending on the widths of the modes.

\subsection{Spatial intensity distribution of QNMs within quasi-1D disordered samples}

In order to fully characterize the QNMs and their relationship to peaks in transmittance in quasi-1D samples, it would be desirable to examine the longitudinal profile of QNMs within the media. Because we do not have access to the interior of the multichannel sample, however, we explore the spatial profile of QNMs using numerical simulations based on the recursive Green’s function technique. The Green’s function between points on the incident plane $\mathbf{r}_0$ and within the sample $\mathbf{r}^\prime$ can be expressed in a manner similar to Eq.~(\ref{Azi1}) as a sum of contribution from each of the modes, We find in the simulations that the spatial distribution of the $m$th mode obtained by decomposing the field into QNMs depends weakly upon the excitation point $\mathbf{r}_0$. We therefore average the spatial profile for each QNM over the profiles obtained for all excitation points on the input of the sample. 

\begin{figure}[tbh]
\centering \scalebox{0.5}{\includegraphics{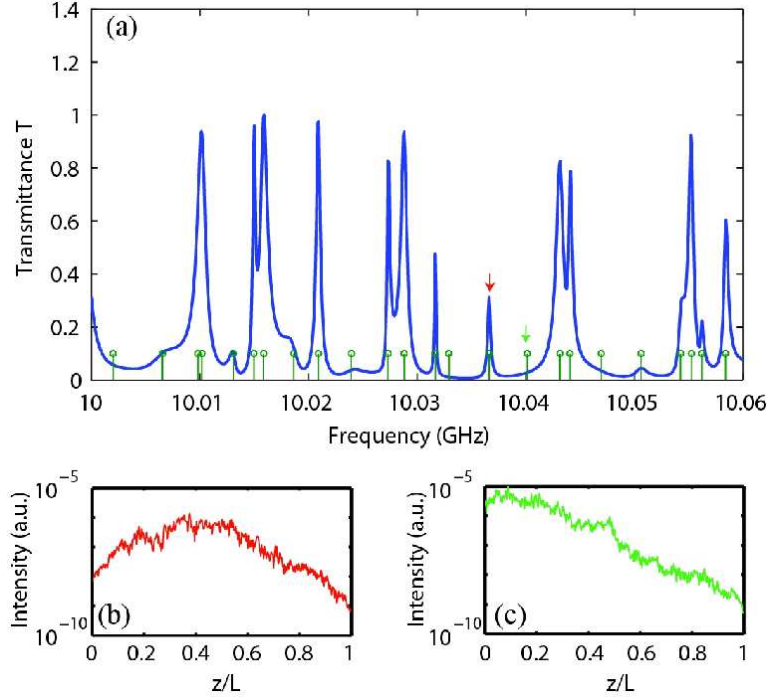}}
\caption{Transmittance spectrum and intensity distribution of QNMs in Q1D disordered samples. (a) Spectrum of transmittance $T$ for a localized sample drawn from an ensemble with $g = 0.26$. The green circles indicate the central frequencies of the QNMs found from a modal decomposition of the field. Longitudinal intensity distributions of two modes with central frequencies indicated by the arrows in (a) are shown in (b) and (c) on a semi-log scale. The average intensity shown is integrated over the transverse dimension of the sample.  The spatial profile of the mode with a peak in transmittance is seen to have a peak in the average intensity in the interior of the sample, while the intensity falls into the sample for a hidden mode. This resembles the behavior of hidden modes found in 1D samples shown in Fig.~\ref{FigIntro}.  
\label{FigEEE}}
\end{figure}

We consider propagation in a sample drawn from an ensemble with a value of $g$ which is below unity but still not too small. In this case, QNMs still overlap but it is yet possible to analyze the field into QNMs. We present in Fig.~\ref{FigEEE} that a spectrum of transmittance in a sample configuration chosen from an ensemble with $g = 0.26$ and $\langle M\rangle = 1.16$, together with profiles of a ordinary and a hidden mode within the spectrum. The nature of propagation in the sample might not differ appreciably from propagation in 1D samples, for which $M = 1$. We find that the intensity distributions integrated over the transverse direction of the hidden mode in the transmission spectrum of the quasi-1D samples falls monotonically within the sample, while the ordinary mode associated with peaks in transmission is peaked in the middle of the sample. 

\section{Analytical calculations of $N_{\mathrm{res}}/N_{\mathrm{mod}}$}

To calculate the average number of TRs in the limit $s\ll 1$, we use the
single-scattering approximation and write the total reflection coefficient $%
r(k)$ of a 1D system as: 
\begin{equation}
r(k)=\Sigma _{n=1}^{N}r_{n}\exp(2ikx_{n}),  \label{eq3}
\end{equation}%
where $x_{n}$ is the coordinate of the $n$-th scatterer. The values $k_{\max
}$, at which the transmission coefficients, $T(k)=1-|r(k)|^{2}$,  has a
local extrema, are defined as the zeros of the function $f(k)\equiv
d|r(k)|^{2}/dk=2\mathrm{Re}\left[ r(k)dr^{\ast }(k)/dk\right] $: 
\begin{equation}
f(k_{\max })=4\mathrm{Im}\Sigma _{n=1}^{N}\Sigma
_{m=1}^{N}r_{n}r_{m}x_{m}e^{2ik_{\max }\left( x_{n}-x_{m}\right) }=0.\label{eq4}
\end{equation}

Assuming first that $\delta d_{i}=0$, we obtain 
\begin{eqnarray} 
f(k)\propto\Sigma _{n=1}^{N}\Sigma _{m=1}^{N}r_{n}r_{m}(m-n)\sin \left[
2k(m-n)d_{0}\right]  \nonumber \\
=\Sigma _{l=1}^{N}\sin \left( 2kld_{0}\right) \left\{ \Sigma
_{n=1}^{N-l}r_{n+l}r_{n}l\right.  \nonumber \\
\left. +\Sigma _{n=l}^{N}r_{n-l}r_{n}l\right\}\equiv\Sigma _{l=1}^{N}\sin
\left( 2kld_{0}\right) a_{l}. \label{eq5}
\end{eqnarray}%
Equation~(\ref{eq5}) is the trigonometric sum $\Sigma _{l=1}^{N}a_{l}\sin \left(
\nu _{l}k\right) $ with ``frequencies'' $\nu _{l}=2ld_{0}$ and random coefficients $a_{l}$. The statistics of the
zeroes of random polynomials have been studied in \cite{Edelman}, where it
is shown that the statistically-averaged number of real roots $N_{\mathrm{%
root}}$ of such sum  at a certain interval $\Delta k$ is 
\begin{equation}
N_{\mathrm{root}}=\frac{\Delta k}{\pi }\sqrt{\frac{\Sigma \nu _{l}^{2}\sigma
_{l}^{2}}{\Sigma \sigma _{l}^{2}}},  \label{eq7}
\end{equation}%
where $\sigma _{l}^{2}=\mathrm{Var}(a_{l})$ is the variance of the
coefficients $a_{L}=\Sigma _{n=1}^{N-l}r_{n+l}r_{n}l+\Sigma
_{n=l}^{N}r_{n-l}r_{n}l$. When the reflection coefficients are uncorrelated, then 
\begin{equation}
\mathrm{Var}(a_{l})= 2(N-l)l^{2}\left(\sigma _{0}^{4}+2\bar{r}^2\sigma_0^2\right),  \label{eq9}
\end{equation}%
where $\sigma _{0}^{2}=\mathrm{Var}(r)$ and $\bar{r}$ is the mean value of $r_i$. The sums in Eq.~(\ref{eq7}) can be
calculated using Eq.~(\ref{eq9}), which yields \cite{GR}: 
\begin{eqnarray}
\Sigma _{l=1}^{N}\sigma _{l}^{2} &=&2\left(\sigma _{0}^{4}+2\bar{r}^2\sigma_0^2\right)\Sigma
_{l=1}^{N}l^{2}(N-l)\simeq \frac{1}{6}\left(\sigma _{0}^{4}+2\bar{r}^2\sigma_0^2\right)N^{4},  \nonumber \\
\Sigma _{l=1}^{N}\nu _{l}^{2}\sigma _{l}^{2} &=&8d_{0}^{2}\Sigma
_{l=1}^{N}\left(\sigma _{0}^{4}+2\bar{r}^2\sigma_0^2\right)l^{4}(N-l)\simeq \frac{4}{15}d_{0}^{2}N^{6}\left(\sigma _{0}^{4}+2\bar{r}^2\sigma_0^2\right).  \label{eq10}
\end{eqnarray}%
From Eqs.~(\ref{eq7}) and (\ref{eq10}) we obtain 
\begin{equation}
N_{\mathrm{root}}=\frac{2\Delta kNd_{0}}{\pi }\sqrt{\frac{2}{5}}=2\frac{%
\Delta kL}{\pi }\sqrt{\frac{2}{5}},\label{eq12}  
\end{equation}%
where $L=Nd_{0}$. Since the number of minima of the reflection coefficient
is equal to the number of TRs, $N_{\mathrm{res}}=N_{\mathrm{root}}/2$, and
the number $N_{\mathrm{mod}}$ of QNMs in the same interval $\Delta k$ is $N_{%
\mathrm{mod}}=\Delta kL/\pi $, from Eq.~(\ref{eq12}) it follows that
\begin{equation}
N_{\mathrm{res}}/N_{\mathrm{mod}}=\sqrt{2/5}.\label{eq13}
\end{equation}

Although this relation was derived for systems with random
reflection coefficients and constant distances between the scatterers, it
also holds for  samples in which these distances  are random  ($\delta d_{i}\neq 0$). In this case, the frequencies $\nu =2ld_{d}$
in Eq.~(\ref{eq5}) should be replaced by $\nu =2|x_{m}-x_{m\pm l}|$. Since
the main contribution to the sums in Eq.~(\ref{eq7}) is given by the terms
with large $l\sim N$, the mean value of $|x_{m}-x_{m\pm l}|$ can be replaced
by $ld_{0}$, in the case of a homogeneous distribution of the distances $d_{n}$ along the system. This ultimately leads to the same result Eq.~(\ref{eq13}).

\section{Hidden modes: simple model}

In Sec. II, QNMs were introduced as solutions of the wave equation satisfying the outgoing boundary conditions.
Their eigenvalues of QNMs, $k^{(\rm mod)}=k^\prime-ik^{\prime\prime}$, can be calculated as roots of the equation $M_{22}=0$, where $\hat{M}$ is the transfer matrix, which connects waves' amplitudes at the left and right sides of the whole system. The transfer matrix of the system which consists of $N+1$ scatterers separated by $N$ intervals has the form:
\begin{equation}\label{eqI}
\hat{M}=\hat{T}_{N+1}\hat{S}_{N}\hat{T}_{N}\hat{S}_{N-1}\cdots\hat{S}_2\hat{T}_2\hat{S}_1\hat{T}_1.
\end{equation}
Here
\begin{equation}
\hat{S}_i=\left\Vert 
\begin{array}{cc}
e^{ikd_i} & 0 \\ 
0 & e^{-ikd_i}%
\end{array}\right\Vert,
\label{eqJ}
\end{equation}
and $\hat{T}_i$ is the transfer matrix through the $i$th scatterer. Assuming that reflection and transmission coefficients are real, $\hat{T}_i$ can be presented as  
\begin{equation}
\hat{T}_i=\left\Vert 
\begin{array}{cc}
1/t_i & -r_i/t_i \\ 
-r_i/t_i & 1/t_i%
\end{array}\right\Vert=\frac{1}{t_i}\left(\hat{I}-r_i\hat{\sigma}_2\right),
\label{eqK}
\end{equation}
where $\hat{I}$ is the unit matrix, and $\hat{\sigma}_2$ is the Pauli matrix.

Omitting denominator $\prod_{i=1}^N t_i$, matrix $\hat{M}$ can be written as ordered product 
\begin{equation}\label{eqL}
\hat{M}=\prod\left(\hat{I}-sr_i\hat{\sigma}_2\right)\hat{S_i},
\end{equation}
where substitution $r_i\to sr_i$ is used. Eq.~(\ref{eqL}) allows  presenting transfer matrix as a power series in $s\ll 1$:
\begin{equation}\label{eqM}
\hat{M}=\prod_{n=1}^N\hat{S}_i+\sum_{n=1}^{N+1} s^n\hat{A}_n,
\end{equation}
where matrix $\hat{A}_n$ contain various ordered products of matrices $\hat{S}_i$ and $n$ Pauli matrices. Matrices $\hat{S}_i$ are diagonal, whereas Pauli matrix is anti-diagonal,  so only even combinations of Pauli matrices contribute to $M_{22}$. Thus,
\begin{equation}\label{eqN}
M_{22}=e^{-ikL}+\sum_{m=1}^{[(N+1)/2]} s^{2m}a_m,
\end{equation}
where $L$ is the sample length, and the coefficients $a_m$ contain various combinations of products of $2m$ reflection coefficients $r_i$ with exponential multipliers $\exp(-ikL+2ik d_{i,j})$, where $d_{i,j}$ are the distances between any ordered pairs of scatterers.

\begin{figure}[tbh]
\centering \scalebox{0.45}{\includegraphics{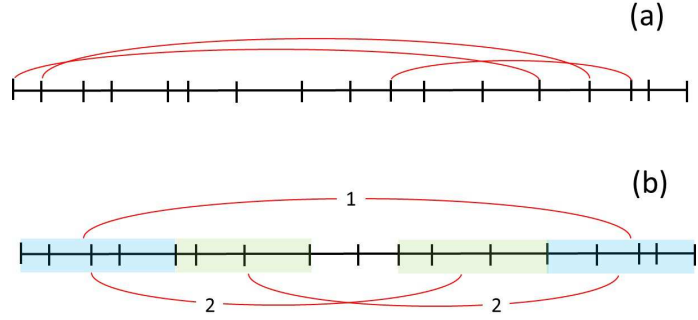}}
\caption{(a) -- coefficients $c_{ij}$ are formed by various pairs of scatterers; (b) -- coefficient $\tilde{c}_1$ contains all possible pairs of scatterers, linked schematically by red line \#1, coefficient $\tilde{c}_2$ is formed by the scatterers from blue and green regions, connected by red lines \#2.}
\label{FigX}
\end{figure}

Neglecting terms with higher than $s^2$ powers in  Eq.~(\ref{eqN}), the dispersion equation, which defines eigenvalues $k$, can be presented as follows:
\begin{equation}\label{eqP}
s^2\sum_{i,j}c_{ij}e^{2k^{\prime\prime} d_{i,j}}=-1,
\end{equation}
where the coefficients $c_{ij}=r_ir_{j}e^{2ik^\prime d_{i,i}}$ are formed by various pairs of the scatterers, as it is schematically shown in Fig.~\ref{FigX}a.
The greater is the distance $d_{i,j}$ between the scatterers, the larger are  exponents $e^{2k^{\prime\prime} d_{i,j}}$ in Eq.~(\ref{eqP}). 

The largest exponents are associated with the pairs of scatterers placed near the opposite ends of the sample. When $k^{\prime\prime}d_0\ll 1$, there are many such pairs, located in blue regions in Fig.~\ref{FigX}b, whose associated exponents are of the same order of magnitude, $e^{2k^{\prime\prime}L}$. Let us combine all such pairs in Eq.~(\ref{eqP}) in one term $\tilde{c}_1$ and characterize them  by one common exponent $e^{2k^{\prime\prime}L}$. The number of scatterers near the sample ends, which form  this group, can be estimated as $n_{\rm eff}\simeq (k^{\prime\prime}d_0)^{-1}\gg 1$, so that the lengths of blue regions in  Fig.~\ref{FigX}b are $\sim n_{\rm eff}d_0\ll L$.

The next group, $\tilde{c}_2$, which is associated with the exponent  of the order of $e^{2k^{\prime\prime}(L-n_{\rm eff}d_0)}$, consists of pairs of scatterers, one from green and another from blue regions in Fig.~\ref{FigX}b. In such a way, Eq.~(\ref{eqP}) can be approximately presented as 
\begin{equation}\label{eqS}
s^2\left(\tilde{c}_1e^{2k^{\prime\prime} L}+\tilde{c}_2e^{2k^{\prime\prime}(L-n_{\rm eff}d_0)}+\ldots\right)=-1,
\end{equation}
Strictly speaking, the phenomenologically introduced number  $n_{\rm eff}$ varies from group to group, but when $k^{\prime\prime}d_0\ll 1$, $n_{\rm eff}$ is large enough and it is possible to neglect its variation. 

The coefficients $\tilde{c}_n$ in Eq.~(\ref{eqS}) are the sums of $n_{\rm eff}$ random vectors in complex plane. For any given sample the lengths of these vectors are fixed, whereas the phases varies from mode to mode, so that the magnitudes of the coefficients $\tilde{c}_n$, been averaged over many modes, can be estimates as
\begin{equation}\label{eqT}
\langle |\tilde{c}_n|\rangle\simeq \sqrt{\langle r^2\rangle^2 n_{\rm eff}}\simeq\langle{r^2}\rangle\sqrt{n_{\rm eff}}.
\end{equation}

Using Eqs.~(\ref{eqS}) and (\ref{eqT}), one can calculate value of $k^{\prime\prime}$, averaged over many modes. When $s^2\ll1$, $\langle k^{\prime\prime}(s)\rangle$ is large and the second term in the parentheses in Eq.~(\ref{eqS}) is small as compared with the first one ($e^{-2k^{\prime\prime}n_{\rm eff}d_0}\ll 1$) and can be omitted. Then, the average solution $\langle k^{\prime\prime}(s)\rangle$ of Eq.~(\ref{eqS}) is
\begin{equation}\label{eqQ}
\langle k^{\prime\prime}(s)\rangle\simeq \frac{1}{2L}\ln\frac{1}{s^2\bar{ r^2}\sqrt{n_{\rm eff}}}=\frac{1}{2L}\left(\ln\frac{1}{s^2\bar{ r^2}}-\frac{1}{2}\ln n_{\rm eff}\right)\simeq \frac{1}{2L}\ln\frac{1}{s^2\bar{r^2}}.
\end{equation}
The dependence $\langle k^{\prime\prime}(s)\rangle$ described by Eq.~(\ref{eqQ}) agrees well with the result of numerical simulations, presented in Fig.~\ref{FigB} by red line.

Expression Eq.~(\ref{eqQ}) describes averaged over many modes dependence $k^{\prime\prime}(s)$, but for any given mode this dependence can be different. Indeed, since $n_{\rm eff}\ll N$  (for example, $n_{\rm eff}\simeq 10$ for $s=10^{-8}$ in the numerical simulation presented in Figs. \ref{FigAA} and \ref{FigB})) fluctuation of the values of $|\tilde{c}_n|$ for different eigenmodes can be rather large. In particular,  $|\tilde{c}_1|$ for a certain mode can be much smaller, than $|\tilde{c}_2|$. Presenting Eq.~(\ref{eqS}) in the form
\begin{equation}\label{eqY}
1+\frac{\tilde{c}_2}{\tilde{c}_1}e^{-2k^{\prime\prime}n_{\rm eff}d_0}=-\frac{e^{-2k^{\prime\prime}L}}{\tilde{c}_1s^2},
\end{equation}
it is easy to see that Eq.~(\ref{eqY}) has solution
\begin{equation}\label{eqU}
k^{\prime\prime}\simeq\frac{1}{2n_{eff} d_0}\ln\left|\frac{\tilde{c}_2}{\tilde{c}_1}\right|,
\end{equation}
when $s^2$ exceeds some critical value $s^2_{\rm crit}$,
\begin{equation}\label{eqZ}
s^2\gg s^2_{\rm crit}=\left|\tilde{c}_1\right|^{-1}\exp\left(-\frac{L}{n_{\rm eff}d_0}\ln\left|\frac{\tilde{c}_2}{\tilde{c}_1}\right|\right).
\end{equation}
Solution Eq.~(\ref{eqU}) is independent of $s$ and represents the hidden QNM (see Fig.~\ref{FigB}).

Recall that $\tilde{c}_n$ are formed by different groups of the reflection coefficients. In general, the similar, independent of $s$, solutions of the dispersion equation appear when magnitudes some first coefficients $\tilde{c}_n$ in Eq.~(\ref{eqS}) are small as compared with magnitudes of the next coefficients.

In order to demonstrate that independent on $s$ solutions of the dispersion equation indeed correspond to the hidden modes, let us consider the system composed of three scatterers only. The dispersion equation Eq.~(\ref{eqS}) for this system is
\begin{equation}\label{eqC}
s^2\left[r_1r_3e^{2ik(d_1+d_2)}+r_2r_3e^{2ikd_2}+r_1r_2e^{2ikd_1}\right]=-1.
\end{equation}
When all $r_i$ are of the same order of magnitude, $r_i\sim r$, and $s$ is so small [$k^{\prime\prime}(s)$ is so large] that $\exp[k^{\prime\prime}(s)d_{1,2}]\gg 1$, the solution $k^{\prime\prime}$ of Eq.~(\ref{eqC}) is 
\begin{equation}\label{eqB}
k^{\prime\prime}\simeq \frac{1}{2(d_1+d_2)}\ln\frac{1}{s^2r^2}.
\end{equation}  
Equations (\ref{eqC}) and (\ref{eqB}) are particular cases of the general formulas (\ref{eqS}) and (\ref{eqQ}). 

If, for example, $|r_1|$ is small as compared with $|r_{2,3}|$, but $s^2\gg s_{\rm crit}^2=|r_1r_3|^{-1}\exp\left[-\frac{d_1+d_2}{d_1}\ln\left|\frac{r_2}{r_1}\right|\right]$,
there is another solution of Eq.~(\ref{eqC}):
\begin{equation}\label{eqG}
k^{\prime\prime}=\frac{1}{2d_1}\ln\left|\frac{r_2}{r_1}\right|.
\end{equation}
This solution is independent on $s$, similarly to the solution Eq.~(\ref{eqU}).

\begin{figure}[tbh]
\centering \scalebox{0.5}{\includegraphics{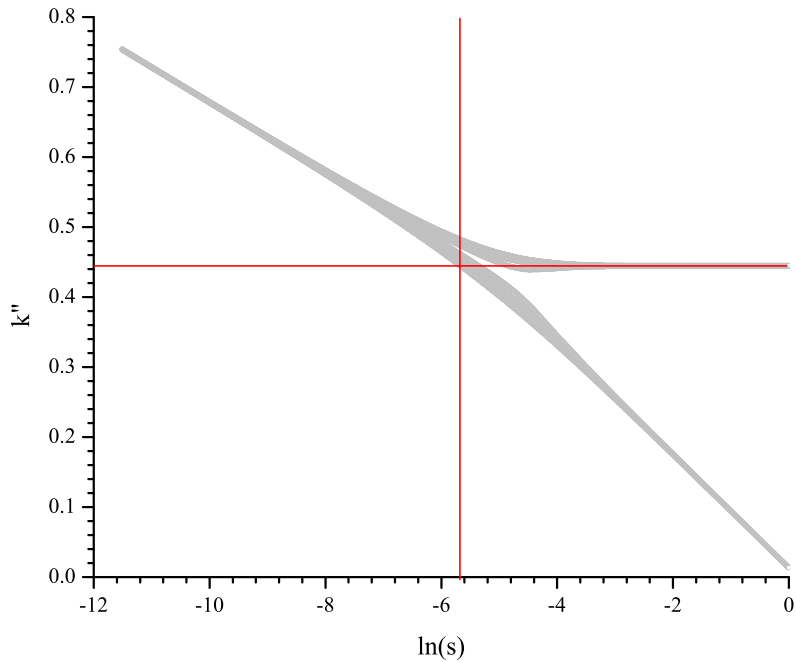}}
\caption{Dependence $k^{\prime\prime}(s)$. Vertical and horizontal red lines mark $s_{\rm crit}$  and $k^{\prime\prime}$ defined by Eq.~(\ref{eqG})}.
\label{FigI}
\end{figure}

\begin{figure}[tbh]
\centering \scalebox{0.5}{\includegraphics{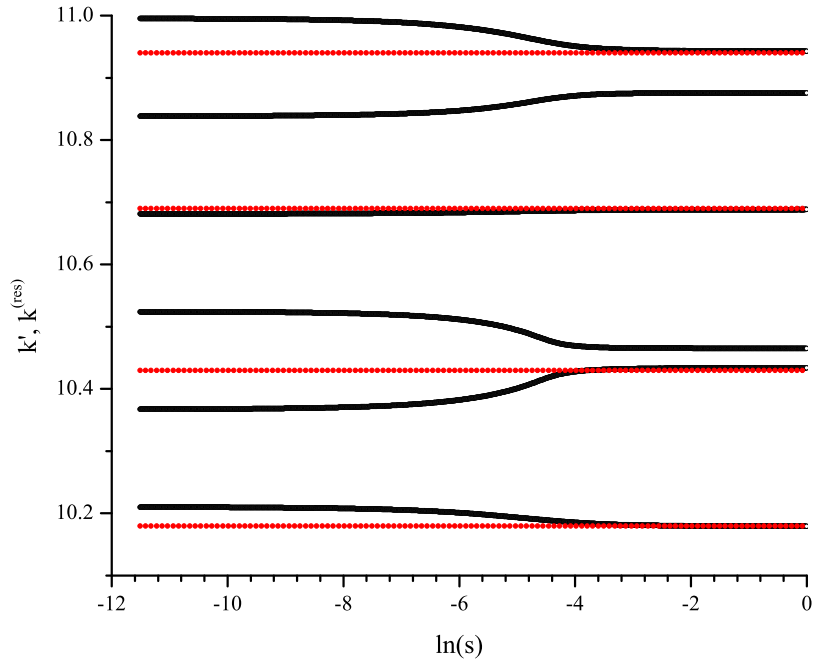}}
\caption{$k^\prime(s)$ (black dots) and $k_{\rm res}(s)$ (red dots). When $s\ll s_{\rm crit}$, the value of $k^{\prime\prime}$ is the same for hidden and ordinary modes.  Nevertheless, hidden modes are invisible in the transmission spectrum. }
\label{FigJ}
\end{figure}

Result of numerical solution of Eq.~(\ref{eqC}) is shown in Fig.~\ref{FigI}. 
Fig.~\ref{FigJ} demonstrates relation between real part of the QNMs' eigenvalues $k^\prime(s)$ and position of the peak $k^{(\rm res)}(s)$ in the transmission spectrum.
Note that hidden modes are invisible in the transmission spectrum even when $s\ll s_{\rm crit}$.
 
\section{Superradiance and resonance trapping in 1D random systems}

The model introduced in the previous Section can be used to study the segregation of superradiant states and trapped modes in regular quantum-mechanical and wave structures and to illuminate the analogy between this phenomenon and existence of two types of QNMs (hidden and ordinary) in disordered systems considered above. Behavior of modes in regular open structures as the coupling to an environment is altered, has been intensively studied in condensed matter physics, optics, and nuclear, atomic, and microwave physics. Common to all these studies is the appearance of two time scales when the coupling to the environment via open decay channels increases and QNMs begin to overlap \cite{new A,new B,new C,new D,new E,new F}; for a review, see \cite{new G} and references therein. When the coupling to the environment is weak, the lifetimes of all states tend to decrease as the coupling increases. As the coupling reaches a critical value, a restructuring of the spectrum of QNMs occurs leading to segregation of the imaginary parts of the complex eigenvalues and of the decay widths. The states separate into short-lived (superradiant) and
long-lived (trapped) states.  This phenomenon is general and, by analogy to quantum optics \cite{new H} and atomic physics \cite{new J,new I,new i}, is known as the superradiance transition. In more complicated structures, such of those consisting of two coupled oscillating subsystems, one with a low and the other with a much higher density of states, the superradiance transition is closely related to the existence of doorway states \cite{new D,new E} that strongly couple to short-lived QNMs with external decay channels. 

It is important to stress that along with the pronounced similarities  between the resonance trapping in many-particle quantum systems, open microwave cavities, etc., and between the ``hidding'' of some of quasi-normal modes in disordered samples there are substantial differences as well. In particular, resonance trapping happens in regular systems considered in \cite{new C,new G} when the coupling of the large number of QNMs to a much smaller number of common decay channels increases. Without disorder, the samples that we consider are perfectly coupled to the environment (total transmission at all frequencies). Finite coupling appears due to disorder, as the result of the interference of multiply-scattered random fields, and the role of the coupling parameter is played by the strength of the scattering inside the system. 

To reproduce the superradince  phenomena in disordered structures we  modify the model slightly by placing the random sample between two reflectors with reflection coefficients $r_L$ and $r_R$, located at distances $\delta_L$ and $\delta_R$ from the edge scatterers. For simplicity, we assume that $\delta_R=\delta_L=\delta$. 
These reflectors can be included in the dispersion equation Eq.~(\ref{eqS}) as additional scatterers as follows:
\begin{equation}\label{eqV}
s^2\left(s^{-2}{r_Lr_R}e^{2ik^{(\rm mod)}(L+2\delta)}+s^{-1} \tilde{c}_0e^{2k^{\prime\prime}(L+\delta)}+\tilde{c}_1e^{2k^{\prime\prime} L}+\tilde{c}_2e^{2k^{\prime\prime}(L-n_{\rm eff} d_0)}+\ldots\right)=-1.
\end{equation}
Here $\tilde{c}_0\propto r_{L,R}$ contains the products $r_{L,R}r_i$ with corresponding exponential multipliers, the largest of which, $\exp\left[2k^{\prime\prime}(L+\delta)\right]$, is separated in the explicit form.

To approach the conditions at which superradiance and resonance trapping occur, we consider below (in contrast to the previous sections) the  evolution of the eigenvalues of a given sample \textit{with fixed} $s$ when $r_{R,L}\to 0$. 

\begin{figure}[tbh]
\centering \scalebox{0.6}{\includegraphics{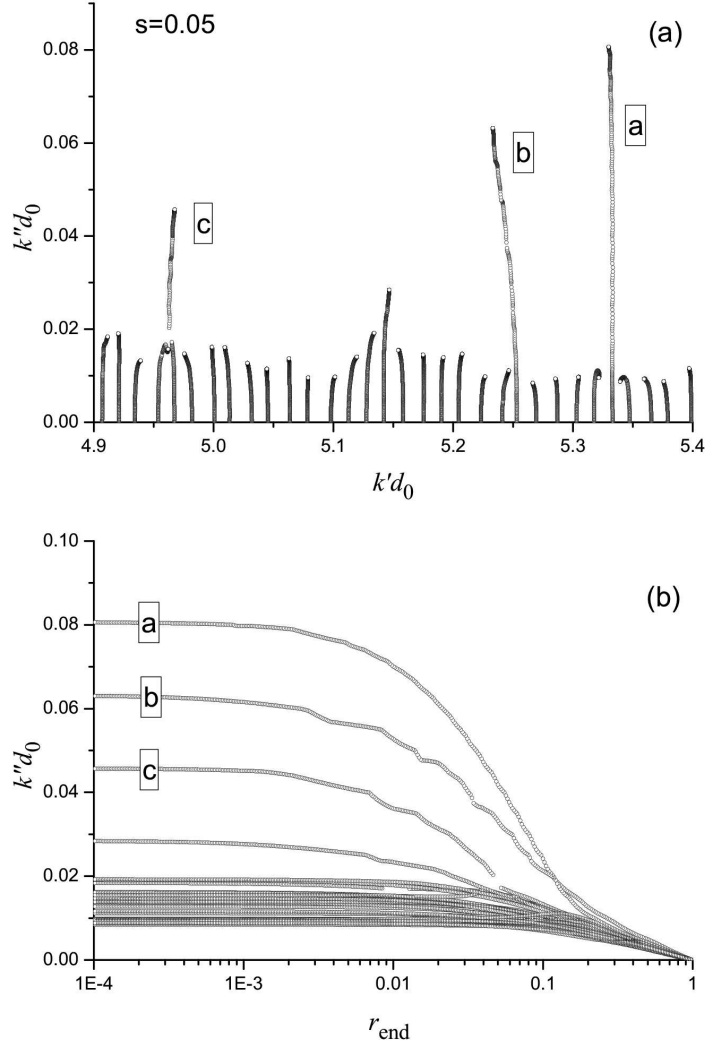}}
\caption{ Two reflectors are placed at the sample ends, $r_L=r_R=r_{\rm end}$. Modes marked by letters  correspond to the same  modes in Figs.~\ref{FigAA}, \ref{FigB}. (a) -- Trajectories of eigenvalues as the coupling grows. (b) -- $k^{\prime\prime}(r_{\rm end})$. The life time of the hidden QNMs decreases much faster than the life time of the ordinary ones.  }.
\label{FigK}
\end{figure}

When the product $|r_Lr_R|$ is large, the first term in the parentheses dominates and the solution of Eq.~(\ref{eqV}) is
\begin{equation}\label{eqW} 
k^{\prime\prime}=\frac{1}{2(L+2\delta)}\ln\left|\frac{1}{r_Lr_R}\right|.\end{equation}
If $\delta=0$, the magnitudes of the exponents in the first three terms are equal. When $|r_Lr_R|\to 0$, the magnitudes of the additional two terms decrease and the solutions of Eq.~(\ref{eqV}) tend to their solutions in the original  sample (without end reflectors), as  shown in Fig.~\ref{FigK}.

\begin{figure}[tbh]
\centering \scalebox{0.6}{\includegraphics{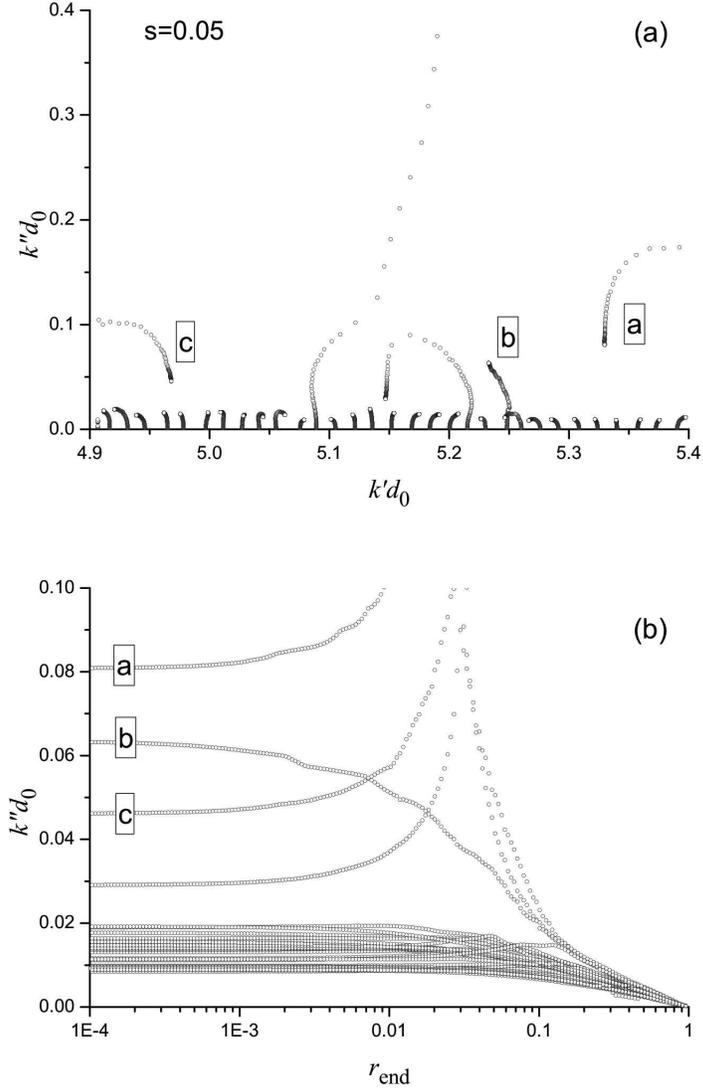}}
\caption{The same as Fig.~\ref{FigL}, but reflectors are placed at a some distance from the sample ends. There is one eigenvalue whose $k^{\prime\prime}$ grows unlimitedly}.
\label{FigL}
\end{figure}

When $\delta\neq 0$, the trajectories of the eigenvalues in the complex plane are more complicate. Although  most of the eigenvalues  finally reach the same positions as in the original sample, there are 
eigenvalues, for which $k^{\prime\prime}\to\infty$ as $r_{L,R}\to 0$ (see Fig.~\ref{FigL}). Indeed,  the first two terms in Eq.~(\ref{eqV}) always dominate  when $\delta\neq0$ and $k^{\prime\prime}\to\infty$. In this case Eq.~(\ref{eqV}) can be written as
\begin{equation}\label{eqZZ}
r_Lr_Re^{2ik(L+2\delta)}+s\left(r_Lr_{N+1}+r_Rr_1\right)e^{2ik(L+\delta)}\simeq 0,
\end{equation}
where the largest term in $\tilde{c}_0$, which corresponds to the largest distance $L+\delta$ between the end reflectors and the sample scatterers, is explicitly presented. Solution of Eq.~(\ref{eqZZ}) 
\begin{equation}\label{eqZZZ}
k^{\prime\prime}=\frac{1}{2\delta}\ln s\left|\frac{r_{N+1}}{r_R}+\frac{r_1}{r_L}\right|
\end{equation}
tends to infinity, when even one of the reflection coefficients $r_{L,R}\to 0$.

The reason why these solutions ``run away'' when the coupling to the environment is maximal, is very simple. The original system without end reflectors has $\Delta kL/\pi$ eigenmodes in the given interval $\Delta k$, whereas the same system surrounded by the reflectors has $\Delta k(L+2\delta)/\pi$ eigenmodes in the same interval. Thus, some of modes should leave this interval $\Delta k$ when the system returns to its original state. 

The superradiant transition in periodic and disordered quantum system, which consist of a sets of potential wells, was studied in \cite{new A} using effective Hamiltonian approach. It was shown that the transition occurs when the  coefficient $\gamma$, which characterizes the coupling with an environment, reaches the value of the coupling $\Omega$ between the wells, $\gamma\simeq\Omega$. In the considered above system $\gamma\simeq 1-|r_{R,L}|^2$ and $\Omega\simeq 1-s^2\langle r^2\rangle$, so that the superradiant transition occurs when $|r_{R,L}|\simeq s\sqrt{\langle r^2\rangle}$. This condition agrees well with presented in Fig.~\ref{FigL}b results.

Hidden modes can be associated with superradiant states, while normal modes are trapped resonances. Thus, $\sqrt{2/5}\simeq 0.63$ correspond to the fraction of the modes which are trapped. This result agrees  with \cite{new C}, where this value was estimated as $>0.58$, and $1-\sqrt{2/5}\simeq 0.37$ is the fraction of the modes which are superradiant. Note, that the original disordered sample is already coupled to the environment, so that the coupling strength is limited by the intrinsic properties of the sample  and cannot exceed this value, even when the end reflectors are fully transparent. 

\section{Conclusions}

In conclusion, we have studied the relationship between spectra of quasi-normal modes
and transmission resonances in open 1D and quasi-1D systems. We start from homogeneous samples, in which each TR is associated with a QNM, and vice versa. As soon as an arbitrarily weak disorder is introduced, this correspondence breaks down: a fraction of the eigenstates becomes hidden, in the sense that the corresponding resonances in transmission disappear. The evolution of the imaginary parts of the eigenfrequencies of the hidden QNMs with changing disorder is also rather unusual. Whereas increasing disorder leads to stronger localization of ordinary modes so that their eigenfrequencies approach the real axis, the imaginary parts of the eigenfrequency of hidden modes changes very slowly (and may even increase when external reflectors are added to the edges) with increasing disorder, and begin to go down only when the disorder becomes strong enough. For weak disorder, the averaged ratio of the number of transmission peaks to the total number of QNMs in a given frequency interval is independent of the type of disorder and deviates only slightly from a  constant, $\sqrt{2/5}$, as the strength of disorder and/or the length of the random sample increase over a wide range. This constant coincides with the value of the ratio $N_{\mathrm{res}}/N_{\mathrm{mod}}$ analytically calculated in the weak single-scattering approximation. As the strength $s$ of disorder keeps growing, ultimately all hidden quasimodes become ordinary. This means that in 1D random systems there exists a pre-localized regime, in which only a fraction of the QNMs are long-lived and provide resonant transmission. If the coupling to the environment is tuned by  external edge reflectors, the superradiace transition can be reproduced. In quasi-1D samples, a genuine diffusive regime exists in which some QNMs coalesce to form a single peak in transmission with width comparable to the typical modal linewidth. In such samples, hidden modes have been discovered experimentally and their proportion of all QNMs in the crossover from diffusion to localization was fairly close to the same constant. The number of peaks in spectra of transmission, as well as in total transmission and in transmittance are nearly the same and fall well below the number of QNMs. Though the ratio $N_{\mathrm{res}}/N_{\mathrm{mod}}$ may be small, we find in microwave experiments and numerical simulations that once the number of QNMs is divided by the effective number of channels contributing to transmission to give $M N_p/N_m$, this function is similar to results in 1D samples. 

\begin{acknowledgments}

We gratefully acknowledge stimulating discussion with K. Bliokh, S. Rotter, R. Berkovits,  J. Page, Hong Chen, and Ping Cheng. We specially thank M. Dennis who drew our attention to paper \cite{Edelman}.

This research is partially supported by the: National Science Foundation (DMR-1207446), RIKEN iTHES Project, MURI Center for Dynamic Magneto-Optics, and a Grant-in-Aid for Scientific Research (S).

\end{acknowledgments}

\end{document}